\documentclass[a4paper,twocolumn,superscriptaddress,longbibliography,pra,nofootinbib]{revtex4-1}

\usepackage{graphicx}
\usepackage{bm}
\usepackage{dsfont}
\usepackage[usenames,dvipsnames]{xcolor}
\usepackage{pstricks}
\usepackage[tight]{subfigure}
\usepackage{verbatim}
\usepackage{units}
\usepackage{natbib}
\usepackage{multirow}
\usepackage{enumitem}
\usepackage{mathrsfs}
\usepackage{leftidx}
\usepackage{xspace}
\usepackage{amsmath,amssymb,amsfonts}

\usepackage[normalem]{ulem}  

\usepackage{hyperref}
\hypersetup{colorlinks=true,linktoc=all,linkcolor=blue,breaklinks=true,citecolor=blue,urlcolor=blue}

\usepackage[english]{babel}
\AtBeginDocument{%
    \newwrite\bibnotes
    \def\bibnotesext{Notes.bib}
    \immediate\openout\bibnotes=\jobname\bibnotesext
    \immediate\write\bibnotes{@CONTROL{REVTEX41Control}}
    \immediate\write\bibnotes{@CONTROL{%
    apsrev41Control,author="
    08",editor="1",pages="1",title="0",year="1"}}
     \if@filesw
     \immediate\write\@auxout{\string\citation{apsrev41Control}}%
    \fi
}%

\newcommand{\kBT}{k_{\text{B}}T}
\newcommand{\rmneq}{{\text{neq}}}
\newcommand{\rmws}{{\text{ws}}}

\newcommand{\beq}{\begin{equation}}
\newcommand{\eeq}{\end{equation}}
\newcommand{\bea}{\begin{eqnarray}}
\newcommand{\eea}{\end{eqnarray}}
\newcommand{\ba}{\begin{align}}
\newcommand{\ea}{\end{align}}

\definecolor{DarkGreen}{rgb}{0,0.7,0}

\begin{document}

\title{Quantifying nonequilibrium thermodynamic operations \\ in a multi-terminal mesoscopic system}

	\author{Fatemeh Hajiloo}
	\affiliation{Department of Microtechnology and Nanoscience (MC2), Chalmers University of Technology, S-412 96 G\"oteborg, Sweden\looseness=-1}

	\author{Rafael S\'anchez}
	\affiliation{Departamento de F\'isica Te\'orica de la Materia Condensada, Condensed~Matter~Physics~Center~(IFIMAC) and Instituto Nicol\'as Cabrera, Universidad Aut\'onoma de Madrid, 28049 Madrid, Spain\looseness=-1}

	\author{Robert S. Whitney}
	\affiliation{Laboratoire de Physique et Mod\'elisation des Milieux Condens\'es, Universit\'e Grenoble Alpes and CNRS, BP 166, 38042 Grenoble, France}
		
	\author{Janine Splettstoesser}
	\affiliation{Department of Microtechnology and Nanoscience (MC2), Chalmers University of Technology, S-412 96 G\"oteborg, Sweden\looseness=-1}

	\date{\today}                                                                                                                                                                                                                                                 
	\begin{abstract}
We investigate a multi-terminal mesoscopic conductor in the quantum Hall regime, subject to temperature and voltage biases. The device can be considered as a nonequilibrium resource acting on a working substance.
We previously showed that cooling and power production can occur in the \textit{absence of energy and particle currents} from a nonequilibrium resource (calling this an N-demon). 
Here we allow energy or particle currents from the nonequilibrium resource, and find that the device \textit{seemingly} operates at a better efficiency than a Carnot engine. To overcome this problem, we define \textit{free-energy} efficiencies which incorporate the fact that a nonequilibrium resource is consumed in addition to heat or power.
These efficiencies are well-behaved for equilibrium and nonequilibrium resources, and have an upper-bound imposed by the laws of thermodynamics.
We optimize power production and cooling in experimentally relevant parameter regimes. 
\end{abstract}

\maketitle

\section{Introduction}

Recently, there has been a lot of interest in the thermodynamics of small-scale electronic devices~\cite{Whitney2018May,Benenti2017Jun,Binder2018}.  The special properties of these devices in the context of thermodynamics stem from quantum and confinement effects. These are prominent in mesoscopic systems such as qubits~\cite{Kosloff2013Jun,Goold2016Feb}, quantum dots~\cite{Sothmann2014Dec}, molecular contacts~\cite{Reddy2007Mar,Dubi2011Mar}, or tunnel junctions~\cite{Pekola2015Feb} whose size is small compared to the coherence length. In most of the cases, the described processes are, analogously to macroscopic thermodynamics, based on the exchange of particles or energy between the system and an equilibrium reservoir. This coupling can be used, e.g., to define heat engines~\cite{Benenti2017Jun}, refrigerators~\cite{Giazotto2006Mar,Courtois2014Jun}, or thermal devices~\cite{Li2012Jul,Martinez-Perez2014Jun,Benenti2016}.

This paper considers ``baths" that can be in a nonequilibrium (or nonthermal) state, without a uniquely defined temperature. This is specific to devices that are smaller than their thermalization length. Such a situation is rare at the macroscopic scale, but is common in nanoscale devices.
It has similarities to recent studies of engines coupled to reservoirs engineered to contain squeezed-states~\cite{Rossnagel2014Jan,Correa2014Feb,Abah2014May,Manzano2016May,Agarwalla2017Sep,Manzano2018Oct,Niedenzu2018Jan,Ghosh2018Mar} or other quantum correlations~\cite{Scully2003Feb,Francica2017Mar}.
Here we identify thermodynamic processes that perform useful tasks by consuming a nonequilibrium distribution \textit{in addition} to consuming heat or work. We have previously shown~\cite{Whitney2016Jan,Sanchez2019Nov} that the consumption of a non-equilibrium distribution alone (without the consumption of heat or work) can produce power or perform cooling. This effect challenges the notion of usual thermal engines. For this reason we called it an N-demon~\cite{Sanchez2019Nov}, where the ``N'' stands for nonequilibrium.

\begin{figure}[b]
\includegraphics[width=\linewidth]{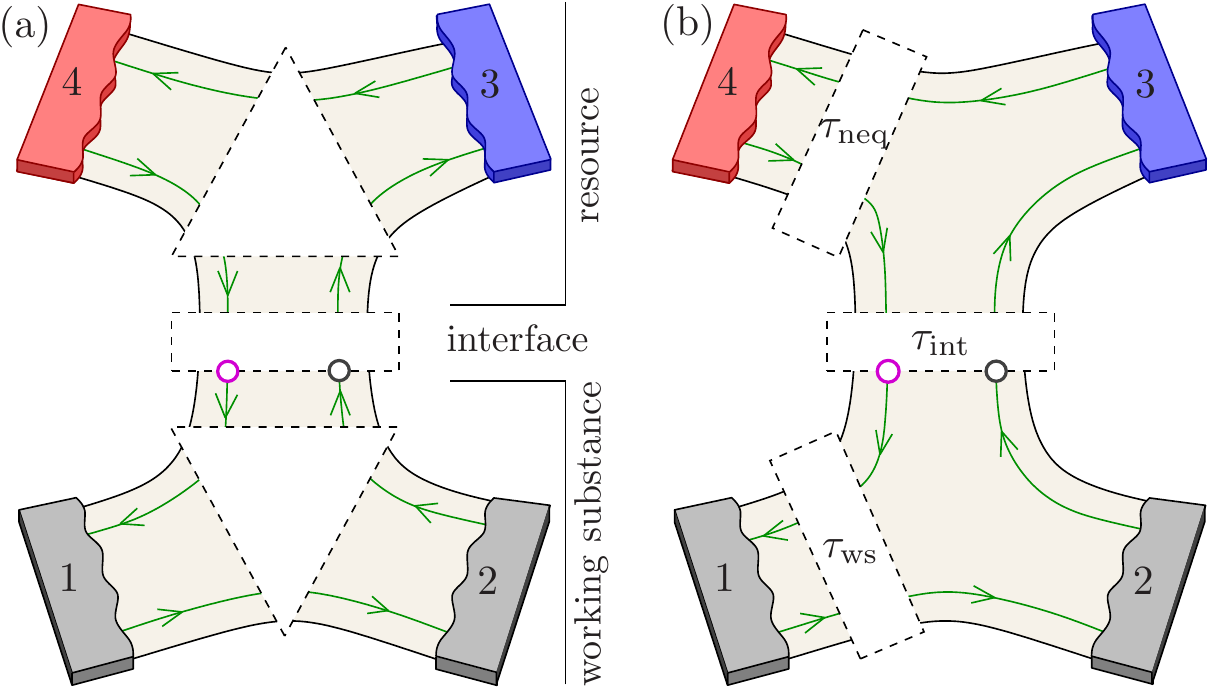}
\caption{Injection of a nonequilibrium distribution into a working substance via quantum Hall edge states. (a) The system is divided into working substance (including terminals 1 and 2), a nonequilibrium resource region (including 3 and 4) and an interface. Each of them contains a scattering region (white areas). (b) Minimal requirements on the position of the scattering regions for power production (or cooling) exploiting a nonequilibrium state as a resource. A nonequilibrium state is injected into the working-substance by mixing the distributions of terminals 2 to 4 at the regions defined by transmission probabilities $\tau_\rmneq$ and $\tau_{\rm int}$. 
The magenta circle indicates where this distribution is injected into the working substance. The black circle indicates the counter-flow.
The injected distribution splits at a scattering region in the working substance, whose transmission $\tau_\rmws$ defines its response.
\label{fig:setup}}
\end{figure}

Here, we extend our previous work~\cite{Sanchez2019Nov}, by relaxing the conditions of \textit{zero} particle and energy flow and investigate the impact of the incoming nonequilibrium distribution on the output power and cooling power of the device. This leads to devices whose operation is partly conventional thermodynamics (i.e.~consuming heat or power)
and partly demonic.
We show that a conventional efficiency can then exceed the Carnot limit, and use this as a sign of the demonic action of the nonequilibrium resource.
While this effect is reminiscent of the action of a Maxwell demon~\cite{HarveyLeff2002Dec}, the resolution of the paradox of the apparent violation of the laws of thermodynamics is different~\cite{Sanchez2019Nov}. 

We wish to compare different machines that operate using demonic resources (nonequilibrium-ness), conventional thermodynamic resources (heat or work), or a combination of the two.  For this we propose \textit{free-energy efficiencies} as a comprehensive measure. We show that these efficiencies are well-behaved for any combination of demonic and conventional thermodynamic operations, and have upper-bounds given by the laws of thermodynamics, see Sec.~\ref{sec:free}.
For example, a machine that produces power $P$ has a free-energy efficiency of the form \begin{eqnarray}
\eta^\text{free}\ \equiv \ P \big/(-\dot{F}_\text{neq}) \ \leq\  1,
\end{eqnarray}
so the power produced can never exceed $-\dot{F}_\text{neq}$.
Here, $-\dot{F}_\text{neq}$ is the rate of reduction of a free energy for a nonequilibrium (or equilibrium) resource. 
This free-energy takes the form of a Helmholtz potential \cite{Callen1985Sep}, 
\begin{eqnarray}
F_\text{neq}=U_\text{neq}-T'S_\text{neq},
\end{eqnarray}
where $U_\text{neq}$ and $S_\text{neq}$ are the nonequilibrium resource's internal energy and entropy. The critical point is to identify which temperature $T'$ appears in this formula; we will argue (see Sec.~\ref{sec:free}) that 
it is the ``ambient'' temperature; i.e. the temperature that all reservoirs would have if there were no mechanisms keeping them hot, cold or out-of-equilibrium.
Sec.~\ref{sec:free} discusses concrete examples, along with the equivalent of the free-energy efficiency for refrigerators.

Such free-energy efficiencies have long been discussed for thermodynamics close to equilibrium (in the linear response regime) for situations as diverse as biological \cite{Caplan1966May} and mesoscopic \cite{Jiang2014Oct} systems.  However, the beauty of free-energies (and efficiencies constructed out of them) is that they also apply arbitrarily far from equilibrium.  This makes them ideal for quantifying the non-equilibrium resources that interest us here.

The class of systems that we study here, shown in Fig.~\ref{fig:setup}, are chosen to be both simple to model and experimentally accessible.  In this context, we make three important choices. Firstly, the nonequilibrium distribution used as a resource is purposefully \textit{made} by mixing the flow from two equilibrium reservoirs. While a nonequilibrium distribution could occur in many different ways, the situation analyzed here should be easily controllable in an experiment and the calculation of the entropy production is particularly insightful. 
Secondly, by tuning reservoir temperature and bias, one can make the device operate as a N-demon, a usual heat-engine (or refrigerator), or a combination of the two.
Thirdly, we consider a multi-terminal electronic realization in the quantum Hall regime. This allows the implementation of complex devices~\cite{Sanchez2015Jul,Hofer2015May,Sanchez2016Jan,Samuelsson2017Jun,Roura-Bas2018Feb,Vannucci2015Aug} such as that in Fig.~\ref{fig:setup}. They have been well studied for nanoscale steady-state heat engines~\cite{Sothmann2014Aug,Sanchez2015Apr,Lopez2014Dec,Sanchez2019Jun}, since they provide excellent control over the electron flow via spatially-separated chiral channels~\cite{Buttiker1992Jan,Granger2009Feb,Nam2013May,Jezouin2013Nov,Sivre2017Oct,Ota2019Feb,Marguerite2019Oct,Marguerite2019Oct,Akiyama2019Dec,Rodriguez2020May,Rosenblatt2020Jun}. 
This way the energy distribution of electrons injected into the working substance at the magenta circle in Fig.~\ref{fig:setup} and of those leaving at the black circle can be measured~\cite{Altimiras2009Oct,LeSueur2010Jul,Altimiras2010Nov,Krahenmann2019Sep}. 

The multi-terminal system shown in Fig.~\ref{fig:setup} can be imagined as consisting in a working substance region, a resource region and an interface where these two regions are put in contact. In the resource region, electrons injected from two or more terminals at possibly different temperatures and electrochemical potentials are scattered and mixed. As a result, outgoing electrons which are injected into the working substance through the interface have a non-thermal distribution. This means they are not described by a Fermi function.  The most striking implication of this ~\cite{Sanchez2019Nov,Whitney2016Jan} is that power production and cooling in the working substance is possible even in the absence of an average particle and energy current between the resource region and working substance.
This is a general statement and should apply in many systems, including optics~\cite{Sanchez2019Nov}. In cases where energy-transfer is done via Coulomb interaction~\cite{Whitney2016Jan,Bhandari2018Jul} rather than particle exchange, the system can be understood as a type of autonomous Maxwell demon~\cite{Sanchez2019Oct}.

Multi-terminal steady-state heat engines are thought to be useful for applications because  (i) they can separate the working substance and resource, thereby limiting heating of the working substance~\cite{Sanchez2011Feb,Thierschmann2015Aug,Sothmann2012May,Roche2015Apr,Mazza2015Jun,Whitney2016May},
and (ii) they can profit from having multiple input resources \cite{Mazza2014Aug,Jiang2014Nov} and generating multiple outputs \cite{Jiang2014Oct,Entin-Wohlman2015Feb,Lu2017Jul,Lu2019Sep}. In the system we analyze here, the situation is even better: no heat transport is required and work production can even be accompanied by a cooling of the working substance. This makes the device interesting for applications in which a nonequilibrium state occurs as the ``waste" of another nanoscale device's operation.

\subsection{Outline of this paper}

This paper is organized as follows. 
Sec.~\ref{sec:setup} presents the quantum Hall setup and the scattering approach. 
Sec.~\ref{sec:noneq} describes the demonic action of the nonequilibrium as a thermodynamics resource for power production and cooling. 
It shows the inadequacies of the usual definitions of efficiency for such nonequilibrium systems. It divides such systems into two classes of demonic action that we call ``strict'' and ``relaxed''. Crucially, Sec.~\ref{sec:free}  defines \textit{free-energy efficiencies}, 
which properly account for the nonequilibrium resources consumed in the thermodynamic process.  It then studies these efficiencies for our quantum Hall device. In Sec.~\ref{sec:thermalize}, we show how these efficiencies change under thermalization of the resource.
Finally, Sec.~\ref{sec:detailed_analysis} discusses physical constraints, and explores experimentally relevant parameter regimes. Our conclusions are given in Sec.~\ref{sec:conclusions}.

\section{Setup and theoretical approach}\label{sec:setup}

\subsection{Basic setup}\label{sec:basic_setup}

The conductor we study is a four-terminal mesoscopic conductor in the quantum Hall regime, subject to a strong perpendicular magnetic field. Electronic charge and energy transport takes place along chiral edge states, indicated by green lines with arrows in Fig.~\ref{fig:setup}.

In many cases, electron transport can well be understood in terms of a single-particle picture, see, e.g., Refs.~\cite{Blanter2000Sep,Ihn,Benenti2017Jun}. Here, this has the advantage that any hidden presence of autonomous feedback~\cite{Strasberg2013Jan,Koski2015Dec,Shiraishi2015Apr,Kutvonen2016Feb,Ptaszynski2018Jan,Sanchez2019Oct} can be excluded as the origin of the discussed demonic effect. 
In this regime, transport in the conductor can be described by a scattering matrix formalism for non-interacting electrons~\cite{Buttiker1992Jan}. 
Then, transport through the device is determined by the transmission probabilities of the scattering regions (represented by white dashed areas in Fig.~\ref{fig:setup}), and by the occupation of the channels incoming from the terminals. These occupations are given by Fermi functions
\begin{equation}
f_\alpha(E)=\left\{1+\exp[(E{-}\mu_\alpha)/k_\text{B}T_\alpha]\right\}^{-1},
\end{equation}
with electrochemical potentials $\mu_\alpha$ and temperatures $T_\alpha$, which can in principle be different in each of the four considered terminals. 
Here, we fix the average potential of terminals 1 and 2 as the zero of energy, namely $\mu_1=\mu_0+\Delta\mu/2\equiv\Delta\mu/2$ and $\mu_2=\mu_0-\Delta\mu/2\equiv-\Delta\mu/2$ with $\mu_0\equiv0$. 
Let us discuss the properties of the three different regions as pictured in Fig.~\ref{fig:setup}(a).

\subsubsection{Resource region} 

The aim of the resource region is to create a nonequilibrium distribution, by mixing different equilibrium ones. This region can hence be interpreted as a \textit{single} terminal that is characterized by a nonequilibrium (or nonthermal) state. 
To create the wanted nonequilibrium distribution, an appropriate scatterer has to be introduced in it. The optimal position of this scatterer depends on the symmetry breaking from the quantum Hall effect that makes the electron motion chiral. Fixing the magnetic field, the minimal requirement to create a nonequilibrium distribution is positioning a scattering region as indicated in the upper third of Fig.~\ref{fig:setup}(b). We characterize this scatterer by the transmission probability $\tau_\text{neq}(E)$. It allows for the mixing of the nonequilibrium distribution. 
In general, an energy-dependence of $\tau_\text{neq}(E)$ is not strictly required, but can be useful to increase the output power or to facilitate the realization of the demonic effect. However, in the linear-response regime, it can be shown that the energy-dependence of $\tau_\text{neq}(E)$, as well as of the other transmission probabilities introduced in the following, can under certain circumstances be  a necessary condition to have a ``demonic action" of this setup.

In Secs.~\ref{sec:noneq} and \ref{sec:detailed_analysis}, we will employ different scatterers as examples. We will consider a quantum point contact (QPC) whose transmission is a step of width $E_1$ at the threshold energy $E_0$~\cite{Buttiker1990Apr}:

\begin{align}
\label{eq:qpcgen}
 \tau_\text{QPC}(E,E_0,E_1)= \left\{1+\exp{[-2\pi(E{-}E_0)/E_1]}\right\}^{-1}.
\end{align}
For the QPC we will assume two limiting cases; one with large $E_1$, so that $\tau_{\rm QPC}$ can be treated as constant, 
and one with small $E_1$, so that the transmission probability can be treated as a sharp step:
\begin{align}
\label{eq:qpc}
 \tau_\text{QPC}(E,E_0)= \Theta(E-E_0).
\end{align}
We will furthermore consider a quantum dot (qd) represented by a resonance at $\epsilon_0$ with broadening $\Gamma$~\cite{Buttiker1988Jan}:
\begin{align}
\label{eq:qd}
\tau_\text{qd}(E,\epsilon_0,\Gamma)=\Gamma^2/[4(E-\epsilon_0)^2+\Gamma^2].
\end{align}
All parameters can be externally tuned as a control knob by means of gate voltages~\cite{parameters}. 
When placed in the resource region, we will call the characteristic energies $E_0=E_\text{neq}$ (for the QPC) or $ \epsilon_0=\epsilon_\text{neq}$ (for the quantum dot). 

\subsubsection{Central region} 
The central region defines the connection point between the resource region and the region where work is done. We choose to place a scatterer with an arbitrary transmission probability $\tau_\text{int}$ in this central region. It allows to fully separate the two regions of interest by pinching off its transmission. At the same time, the presence of a partially open scatterer defines the \textit{point} at which the created nonequilibrium distribution is injected into the working substance. The conceptually most simple interface region for our purposes is an open (or at least energy-independent) scatterer. In this way we avoid that it contributes to an additional mixing of a nonequilibrium distribution, which is external to the resource, or an additional power production, which can not clearly be associated to the working substance. In most of the discussion below we will hence consider $\tau_{\rm int}=1$. 

In Sec.~\ref{sec:probe}, we will \textit{replace} this simple central region by a region with energy-resolved detectors and additional terminals acting as temperature and voltage probes.

\subsubsection{Working substance} 
The injected nonequilibrium distribution is to be used for work production or cooling in the working substance. Therefore an additional scattering region with transmission probability $\tau_\text{ws}$, is needed, where the injected nonequilibrium is exploited, before it can equilibrate in the lower part of the conductor, see the lower left part of Fig.~\ref{fig:setup}(b). Again, we fix its  position based on the choice of the magnetic-field direction~\cite{Sanchez2015Apr}. Also here, as examples for this transmission probability we will later choose a QPC or a quantum dot, given by Eqs.~\eqref{eq:qpc} and \eqref{eq:qd}, this time with $E_0=E_\text{ws}$ or $\epsilon_0=\epsilon_\text{ws}$. These types of junctions have been known for their thermoelectric properties since long ago~\cite{Streda1989Feb,Molenkamp1990Aug,Beenakker1992Oct,Staring1993Apr,Dzurak1993Sep,Svensson2012Mar,Josefsson2018Jul}.
Note however, that even in the working substance, no broken electron-hole symmetry is required for the scatterer in order to allow for power production, as we will show in Sec.~\ref{sec:detailed_analysis}. This is in strong contrast to thermoelectric devices with usual heat baths.

\subsection{Currents and engine output}\label{sec:define_currents}

Having introduced three  scattering matrices, we can write down particle, $I_\alpha\equiv I^{(0)}_\alpha$, and energy currents, $I^E_\alpha\equiv I^{(1)}_\alpha$, flowing into all reservoirs $\alpha$,
\begin{subequations}
\begin{align}
I_\alpha^{(\nu)}&=\frac{1}{h}\int dE \  E^\nu\ j_\alpha(E), \label{eq_Il}
\end{align}
with
\begin{align}
j_1(E)&=\tau_\text{ws}(E)\left[\tilde{f}(E)-f_1(E)\right] \label{eq_I1}
\\
j_2(E)&=f_1(E){-}f_2(E){+}\left[1{-}\tau_\text{ws}(E)\right][\tilde{f}(E){-}f_1(E)]\label{eq_I2}
\\
j_3(E)&=\tau_\text{int}(E)f_2(E)-\{[1{-}\tau_\text{int}(E)]-\tau_\text{neq}\}f_3(E)\nonumber\\
&\quad+[1{-}\tau_\text{int}(E)]\tau_\text{neq}(E)f_4(E) \\
j_4(E)&=\tau_\text{neq}(E)\left[f_3(E)-f_4(E)\right] .
\end{align}
\end{subequations}

Here, the superscript $\nu$ indicates the power with which the energy $E$ enters the integrals. We have introduced the nonequilibrium distribution function injected into the working substance from the central scatterer [i.e., at the magenta circle in Fig.~\ref{fig:setup}(b)],
\begin{align}
\tilde{f}(E)=& \tau_\text{int}(E)\tau_\text{neq}(E)f_4(E)+\tau_\text{int}(E)\left[1-\tau_\text{neq}(E)\right]f_3(E)\nonumber\\
&+\left[1-\tau_\text{int}(E)\right]f_2(E).\label{eq:ftilde}
\end{align}
Already at this general level, we can make some important observations. (i) Only in the special case $\tau_\text{int}(E)\equiv 1$ is the nonequilibrium distribution exclusively composed by distributions from the upper resource region. While in principle, the effects to be discussed here do not rely on where the nonequilibrium distribution is created (this will be shown in Sec.~\ref{sec:detailed_analysis}), we will often choose this special case $\tau_\text{int}(E)\equiv 1$ to get a clear, intuitive understanding of the role of the different subparts of the system. When fixing $\tau_\text{int}(E)\equiv 1$, we furthermore also notice that the nonequilibrium distribution can fully be exploited for work production or cooling. Otherwise, a part of it would be dumped into reservoir 3 without being exploited. (ii)  The charge or heat flow expected from possible voltage and temperature biases in reservoir 2 [see first contributions to Eq.~(\ref{eq_I2})] can be reversed when appropriately choosing the created nonequilibrium distribution and the transmission probability $\tau_\text{ws}(E)$ at the lower scatterer [see the last contribution to Eq.~(\ref{eq_I2})].

With the particle and energy currents written above, we write the heat currents, $J_\alpha$, as
\begin{align}
\label{eq:heatcurr}
    J_\alpha & =I^{E}_\alpha-\mu_\alpha I_\alpha.
\end{align}
These currents constitute the main observables quantifying the output of the steady-state engines implemented here.
Particle and energy conservation ensure that $0=\sum_\alpha I_\alpha=\sum_\alpha I^E_\alpha$, where the sums are over all reservoirs.  This gives the 
first law of thermodynamics
\begin{eqnarray}
\sum_\alpha J_\alpha + \sum_\alpha \mu_\alpha I_\alpha =0
\end{eqnarray}
where the first sum is the heat flow into reservoirs and the second is the power generated (work done per unit time)  by pushing a current against an electrochemical potential. These currents also always obey the second law of thermodynamics,
see e.g. section 6.4 of Ref.~\cite{Benenti2017Jun}.

When we are interested in electrical power generated in the working substance, we define it as
\begin{equation}
\label{eq:PowerHeatEngine}
P=(\mu_1-\mu_2)I_1.
\end{equation}
Ref.~\onlinecite{Whitney2013Mar} gave a simple upper bound for the power generated in a usual (thermoelectric) system with hot and cold heat baths, with temperatures $T_\text{hot}$ and $T_\text{cold}$.
A less simple but tighter upper bound is \cite{Whitney2014Apr,Whitney2015Mar}
\begin{equation}\label{eq:WhitneyPower}
P^{\text{qb}}=\frac{A_0\pi^2k_{\text{B}}^2}{ h}\, (T_{\text{hot}}-T_{\text{cold}})^2,
\end{equation}
where $A_0 \simeq 0.0321$.
On the other hand, when the device is operated as a refrigerator, we are interested in the cooling power, $J_\text{cold}$, given by the heat current flowing out of the reservoir to be cooled. If one is cooling reservoir 1 then  $J_\text{cold}=-J_1$, however if one is cooling reservoir 2 then $J_\text{cold}=-J_2$. In a standard refrigerator, the bound on the cooling power, similar to Eq.~(\ref{eq:WhitneyPower}), is given by~\cite{Pendry1983Jul}
\begin{equation}\label{eq:pendrybound}
    J^{\text{qb}}=\frac{\pi^2k_{\text{B}}^2}{6 h }T_{\text{cold}}^2.
\end{equation}
Results on cooling and power  production in the working substance exploiting the injected nonequilibrium distribution will be presented in detail in Sec.~\ref{sec:noneq}.

\subsection{Conventional efficiencies of power production and cooling}\label{sec:efficiency}
%
In this section we introduce the three conventional efficiencies for three thermodynamic devices; heat engines, standard refrigerators and absorption refrigerators.  In all cases the conventional efficiency is a dimensionless ratio of the device's desired output to the resource expended by the device. For example, if the device is a usual heat engine, the desired output is work (in our case electrical power), and the resource expended is the heat in a hot reservoir. 
The laws of thermodynamics set upper bounds on these efficiencies.
Let us now treat them one by one, in the context of our four-terminal mesoscopic conductor.

In a usual heat engine, the performance can be quantified via a conventional efficiency, given by the ratio between produced power and heat currents extracted from the resource. 
In this case, the resource is assumed to provide no electrical power, so we ensure that \textit{electrical currents} flowing from the top to the bottom region are zero, $I_3{+}I_4{\equiv}~0$. Then the efficiency would correspond to
\begin{equation}\label{eq:effJ}
    \eta^J=\frac{P}{-(J_3+J_4)},
\end{equation}
see Sec.~\ref{sec:engine}. However, an important remark needs to be made here. 
For a nonequilibrium distribution induced by the resource due to a potential bias, $\mu_3-\mu_4$, the electronic \textit{heat currents} into or out of the terminals contains a Joule heating contribution, despite $I_3+I_4=0$. Consequently, the heat current through the interface differs from energy current ---something that is not the case if the energy flow is entirely due to temperature differences. 
Therefore, the correct definition for the efficiency in voltage biased devices with particle exchange is
\begin{align}\label{eq:effE}
    \eta^E=\frac{P}{-(I^E_3+I^E_4)}.
\end{align}
Otherwise, when comparing our device with an engine operated with a  usual heat bath, the heat current, $J_3+J_4$, (including Joule heating) would correspond to an effective bath with a lower temperature.
We will discuss the difference between these definitions for the conventional efficiency in Sec.~\ref{sec:engine}.

The well-known bound on the conventional efficiency of a heat engine in contact with a hot and a cold bath (with temperatures $T_\text{hot}$ and $T_\text{cold}$) is the Carnot efficiency, relying on the laws of thermodynamics,
\begin{equation}\label{eq:effCarnot}
    \eta^{\text{Carnot}}=1-\frac{T_\text{cold}}{T_\text{hot}}.
\end{equation}
However, one cannot straightforwardly associate a temperature with the nonequilibrium resource region. Section~\ref{sec:probe} shows how to do this, by measuring the nonequilibrium distribution with a temperature probe.

For a refrigerator, the conventional efficiency is called the coefficient of performance (COP) and is given by the cooling power divided by the resource absorbed. A refrigerator can either use electrical power as a resource (standard refrigerator) or use heat (absorption refrigerator). We will compare our device to both of these.

In a standard refrigerator the resource injects electrical power $P_\text{in}$ but supplies no heat, from which the first law of thermodynamics tells us that the heat flow out of the working substance equals the power injected, so we have $J_1+J_2=P_\text{in}$. If the refrigerator extracts a heat $J_\text{cold}$ from the cold reservoir (so if this is reservoir 2, then $J_2=-J_\text{cold}$), then
the COP is given by
\begin{align}\label{eq:def_COP_TE}
\text{COP}^\text{}    &  \equiv\frac{J_\text{cold}}{P_\text{in}}=\frac{J_\text{cold}}{-(I_3+I_4)\delta\mu}.
\end{align}
The potential difference $\delta\mu=\mu_\text{p}-\mu_0$ is between the resource region and the working substance at $\mu_0$. However, $\mu_\text{p}$ is not straightforwardly defined for a nonequilibrium resource. We define it as that measured by a voltage probe, as discussed in Sec.~\ref{sec:probe}.
The Carnot upper bound on this COP for cooling induced by electrical work is
\begin{equation}\label{eq:Carnot_COP_TE}
    \text{COP}^\text{Carnot{}}=\frac{T_\text{cold}}{T_\text{hot}-T_\text{cold}}.
\end{equation}

In contrast, an absorption refrigerator extracts heat $J_\text{cold}$ from one cold terminal by using heat from a hot reservoir; this is sometimes called \textit{cooling-by-heating}. The heat from these reservoirs is dumped into a reservoir at ambient temperature. Hence there is no charge current into (or out of) the resource, meaning that charge is conserved in the working substance; $I_1+I_2\equiv -(I_3+I_4)=0$. Then heat flow into the working substance from the resource equals $-(I^E_3+I^E_4)$, and the COP is
\begin{align}\label{eq:def_COP_abs}
\text{COP}^\text{abs}   & =\frac{ J_\text{cold}}{-I^E_3-I^E_4}.
\end{align}
For ambient temperature $T_\text{amb}$, its  Carnot bound is
\begin{equation}\label{eq:Carnot_COP_abs}
    \text{COP}^\text{abs;Carnot}=\frac{1-T_\text{amb}/T_\text{hot}}{T_\text{amb}/T_\text{cold}-1}.
\end{equation}

\subsection{Detectors and probes}\label{sec:probe}

\begin{figure}[b!!]
\includegraphics[width=\linewidth]{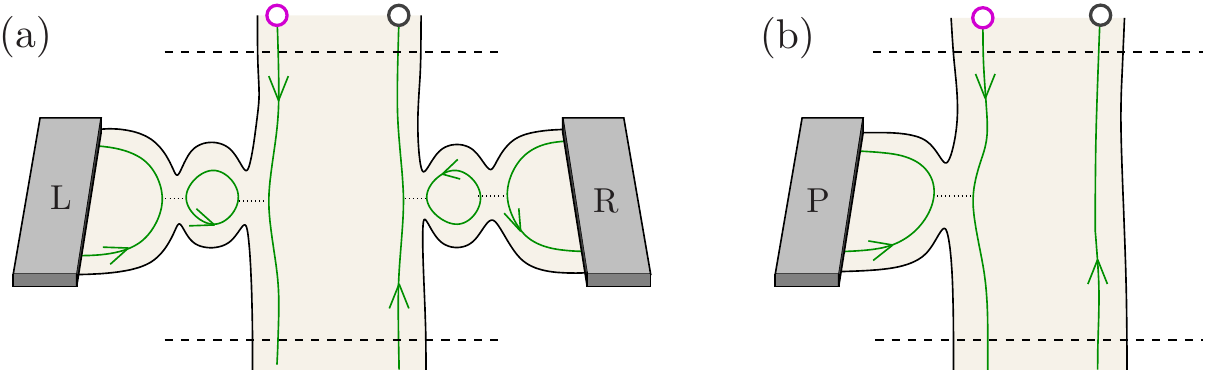}
\caption{(a) Side-coupled quantum dots as detectors for incoming and outgoing distributions. (b) Floating voltage and temperature probe, allowing to (partially) equilibrate the incoming distribution.}
\label{fig:detect}
\end{figure}

To observe the properties of the nonequilibrium distribution, we now complement the setup in Fig.~\ref{fig:setup} by two types of elements. First, we add detectors to read out incoming and outgoing distributions at the central scatterer, see Fig.~\ref{fig:detect}(a). With this, we concretely suggest, how an experimental test of the nonequilibrium character of distributions could be implemented. Second, we add a floating probe contact, see Fig.~\ref{fig:detect}(b). Such a probe can measure an effective temperature and electrochemical potential of the incoming nonequilibrium distribution. We concretely use this model setup to extract the effective temperatures and electrochemical potentials entering efficiency definitions. Furthermore, if desired, the probe can be used to
(partially or fully) equilibrate this nonequilibrium distribution. The elements in Fig.~\ref{fig:detect} are to be added to the central region of Fig.~\ref{fig:setup}.

\subsubsection{Side-coupled detector dots}

To read out the full distribution functions in the central region ---both that flowing from top to bottom, and that flowing from bottom to top---, one can add two side-coupled detector quantum dots to the setup~\cite{LeSueur2010Jul,Altimiras2009Oct} as in  Fig.~\ref{fig:detect}(a). The transmission between the side reservoirs and the main conductor via the quantum dots is taken to have the form $\tau_\alpha(E,\epsilon_0,\Gamma)$ (for $\alpha=$L,R) as in Eq.~(\ref{eq:qd}). 

For the particle current in reservoir L we have
\begin{align}
I_\text{L}(\epsilon_0) =\frac{1}{h}\int dE \ \tau_\text{L}(E,\epsilon_0,\Gamma)\ \left[\tilde{f}(E)-f_\text{L}(E)\right].
\end{align}
If the transmission through the dot is given by a single, sharp peak in energy centred at energy $\epsilon_0$ (which can be varied with a gate) with narrow width $\Gamma$, then
\begin{align}
\tilde{f}(\epsilon_0)-f_\text{L}(\epsilon_0) \,\simeq\,
\frac{h I_\text{L}(\epsilon_0)}{\Gamma}.
\end{align}
If the detector's Fermi distribution, $f_\text{L}(E)$,  is known, one can get the nonequilibrium distribution, $\tilde{f}(\epsilon_0)$, by measuring the current $I_\text{L}(\epsilon_0)$ as a function of $\epsilon_0$. Note that energy resolution with which one can measure the nonequilibrium distribution will be be limited by $\Gamma$.

The distribution of the upwards propagating edge of the conductor [at the black circle in Fig.~\ref{fig:setup}(b)], is measured in the same way by the right side reservoir.

\subsubsection{Floating voltage and temperature probe}

In the previous section, we propose experiments to detect the incoming and outgoing (nonequilibrium) distributions in the considered conductor. To examine the relevance of the nonequilibrium distribution to the physics, it is useful to be able to (partially) force this distribution to equilibrate, without changing its energy or particle number. This can be done by attaching a floating voltage and temperature probe to the conductor, as shown in Fig.~\ref{fig:detect}(b). The probe establishes itself at an electrochemical potential $\mu_\text{p}$ and temperature $T_\text{p}$ which guarantees that no heat and charge currents are flowing into/out of the probe contact (p), i.e.,
\begin{equation}
I_\text{p}^{(\nu)}=\frac{\tau_\text{p}}{h}\int dE\ E^\nu\left[\tilde{f}(E)-f_\text{p}(E)\right]=0.
\end{equation}
Such probes have been used to describe decoherence~\cite{Engquist1981Jul,Buttiker1986Mar,D'Amato1990Apr,Pastawski1991Sep,Sanchez2011Nov,Meair2014Jul},  here we use them to describe the thermalization process.
We have assumed the transmission probability $\tau_\text{p}$ is energy-independent,
to avoid additional thermoelectric effects.  When $\tau_\text{p}=1$, the probe contact injects an equilibrium distribution function $f_\text{p}(E)$ which carries the same amount of charge and energy as the nonequilibrium distribution $\tilde{f}(E)$. Hence, the distribution is the same as if interactions between the electrons in the edge-state had redistributed their energy, turning their nonequilibrium distribution into a thermal one.
For $0<\tau_\text{p}<1$, a partial equilibration of the injected distribution takes place, and it becomes
\begin{equation}
\tilde{f}(E)\rightarrow\tau_\text{p} f_\text{p}(E)+\left(1-\tau_\text{p}\right)\tilde{f}(E)\ .
\end{equation}
Thus one can experimentally extract the role played by the nonequilibrium-ness by tuning $\tau_\text{p}$, see Sec.~\ref{sec:demonic_action}.

Furthermore, this voltage and temperature probe also measures the effective temperature, $T_\text{p}$, and electrochemical potential, $\mu_\text{p}$, which one can associate to the incoming nonequilibrium distribution. We concretely use this to evaluate both the conventional efficiencies and their bounds, as well as the proper free-energy efficiencies, which are determined by the temperatures and potentials of resource region and working substance.  

Such tunnel contacts can also be used as detectors for the local distribution via noise measurements~\cite{gramespacher1999,gabelli2013,Ota2017Apr,Tikhonov2020Jan,Dashti2018Aug}.

\section{Demonic action of a noneqilibrium resource}\label{sec:noneq}

In this section, we demonstrate how a nonequilibrium distribution, injected from the resource region into the working substance, can be exploited to produce electrical power or to cool. We emphasize its properties by first interpreting its performance in terms of conventional thermodynamic efficiencies for systems driven by power or thermal flows, as introduced in Sec.~\ref{sec:efficiency}. This way, we identify two types of regimes, in which the performance of the engine  seemingly exceeds fundamental bounds. We will later name  them the ``strict N-demon" and the ``relaxed N-demon". We refer to these systems as N-demons  for their analogies to Maxwell's demon 
that can  apparently violate the second law, when coupled to a working substance. 

Since we aim to show the principles of the operation in this section, we focus on one simple realization of the four-terminal setup, where both the scatterers in the resource region and the working substance are given by QPCs with sharp step-like transmission probabilities, see Eq.~\eqref{eq:qpc}, as these are expected to give good energy resolution and at the same time to provide large currents~\cite{Kheradsoud2019Aug}. Analytical expressions for this choice of transmission probabilities are given in Appendix~\ref{app:integrals}.

To analyze an experimentally feasible setting, we assume below that the nonequilibrium distribution in the resource region is created at fixed temperatures $T_3$ and $T_4$. Then, the chemical potentials $\mu_3$ and $\mu_4$ can be tuned or can be adjusted to satisfy conditions on the particle and/or energy currents, see for example Eqs.~\eqref{eq:cond_engine}, \eqref{eq:cond_TEfridge} and \eqref{eq:cond_absfridge} below. Note that this is a choice; another choice would be, e.g., to fix the chemical potential and temperature of one terminal and tune those of the other.

\subsection{Steady state power production}\label{sec:engine}

\begin{figure}[t]
\includegraphics[width=\columnwidth]{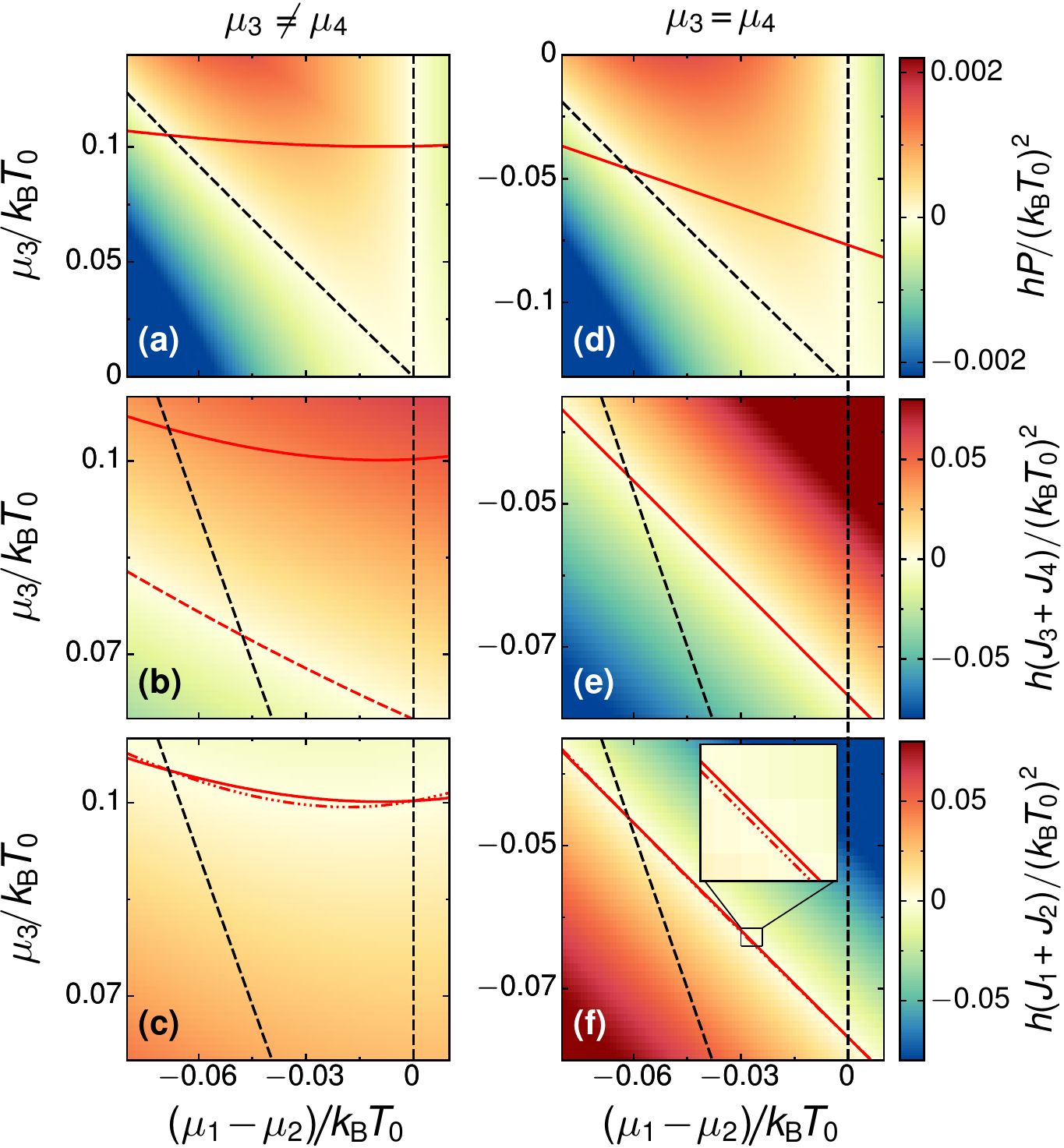}
\caption{ Power production in the absence of a charge flow from the resource region, $I_3{+}I_4=I_1{+}I_2=0$. (a) Produced power, (b) total heat current flowing out of the resource region, $J_3{+}J_4$, and (c) total heat current flowing out of the working substance, $J_1{+}J_2$. For all three panels, we fix $T_1=T_2=T_3= T_0$, but $T_4{=}1.2 T_0$, as well as $\mu_4{=}{-}0.4k_{\text{B}}T_0{\neq}\mu_3$. 
(d)--(f) Equivalent to left column with $\mu_3{=}\mu_4$, temperatures $T_1{=}T_2{=}T_0$, $T_3{=}0.9T_0$ and $T_4{=}1.2T_3$.
In all figure panels, we choose $E_\text{ws}=0$ and fix $E_\text{neq}$ to fulfill Eq.~(\ref{eq:cond_engine}). 
The red solid lines indicate, where we additionally have $I^E_3{+}I^E_4=I^E_1{+}I^E_2=0$ (later identified as a  ``strict N-demon" condition, see Eq.~(\ref{eq:strict_demon})). As a guide for the eye, we show where $J_1{+}J_2=0$ (dashed red line), $J_3{+}J_4=0$ (dashed-dotted red line) and the power $P=0$ (dashed black lines). 
\label{fig:power_production}}
\end{figure}

We study steady-state electrical power production, in analogy to a heat engine, by imposing the condition 
\begin{align}\label{eq:cond_engine}
I_3+I_4&= 0.    
\end{align}
This condition excludes situations where electrical power is injected from the resource region, in order to make the configuration as close to a usual heat engine as possible, see the discussion in Sec.~\ref{sec:efficiency}.

The produced power is given by Eq.~(\ref{eq:PowerHeatEngine}) and the results are shown in Fig.~\ref{fig:power_production}(a) and (d) as a function of $\mu_1-\mu_2$ and $\mu_3$. 
Here, $\mu_1-\mu_2$ is the potential drop against which a particle current is transported in the working substance. Tuning the potential in the resource region, $\mu_3$, influences the nonequilibrium distribution that is injected into the working substance.  To eliminate additional thermoelectric contributions within the working substance, in which we are not interested here, we fix the temperatures $T_1=T_2=T_0$.  We consider both cases where the variation of $\mu_3$ leads to a potential drop in the resource region (left column) and where $\mu_3=\mu_4$ (right column). 
Furthermore, we fix the step energy of the QPC transmission of the working substance to $E_\text{ws}\equiv0$, while $E_\text{neq}$ is adjusted for each value of electrochemical potentials in order to fulfill condition Eq.~(\ref{eq:cond_engine}).  
We find extended regions, where positive power is produced in the working substance. These regions are those between the two dashed black lines in each panel of Fig.~\ref{fig:power_production}. One of the dashed black lines trivially corresponds to $\mu_1=\mu_2$, so the electronic current does no work, and $P=0$. The other dashed black line marks where the current induced by the working substance internal chemical potential bias compensates the one generated by the nonequilibrium resource. The power output hence changes sign when crossing that line, similarly to what occurs at the stall voltage in conventional thermoelectrics.

Importantly, power can even be produced if the energy current across the interface between resource region and working substance equals zero, $I^E_3+I^E_4=I^E_1+I^E_2=0$ i.e., when both particles and energy are conserved in the working substance. This is the effect that was previously predicted in Ref.~\onlinecite{Sanchez2019Nov}, and which we call the strict N-demon below (indicated by the solid red lines in all panels of this and the following figures). Furthermore, we plot the total heat currents flowing out of the resource region and out of the working substance, in panels 
~(b,c) and (e,f). Indeed these panels show that the heat currents change sign within the region of finite power production. 
Note that this sign change of energy/heat currents means that the present engine can do work either by absorbing heat/energy or by releasing it, while the temperature of the terminals is fixed. This is very different from usual engines, where heat is absorbed  from a hot bath to do work (or alternatively a cold bath is used as a resource which is heated up to do work). The origin of this difference lies in the fact that no meaningful temperature can be associated to the nonequilibrium resource, which we exploit here.

However, only in the case without internal bias in the resource region, $\mu_3=\mu_4$, (respectively in the working substance, $\mu_1=\mu_2$), does zero energy current corresponds to zero heat current. The reason for this is that ---even though the particle current across the interface is fixed to zero by Eq.~(\ref{eq:cond_engine})--- the chemical-work contribution to the heat current, $\sum_\alpha\mu_\alpha I_\alpha$, does not vanish, if electrochemical potentials differ from each other. 
In the regions between the solid and the dashed red lines in panels~(b) [respectively between the solid and the dashed-dotted red lines in panel (c)], the energy currents and heat currents out of the resource region (respectively out of the working substance) have opposite signs. 
The internal bias in the resource region can lead to Joule heating
which means that heat is not conserved in the resource region. Thus we cannot assume that the heat flow through the interface is equal to $-(J_3+J_4)$.
Instead we can argue that, in analogy to Eq.~\eqref{eq:heatcurr}, the heat flow through the interface is given by $-[I^E_3+I^E_4 - \mu_\textrm{p}(I_3+I_4)]$, where $\mu_\textrm{p}$ is the effective potential of the resource region (as measured by a voltage probe, as explained in Sec.~\ref{sec:probe}).  However as we take $I_3+I_4=0$, this means that the heat flow through the interface into the working substance is $-(I^E_3+I^E_4)$.

\begin{figure}[t]
\includegraphics[width=\columnwidth]{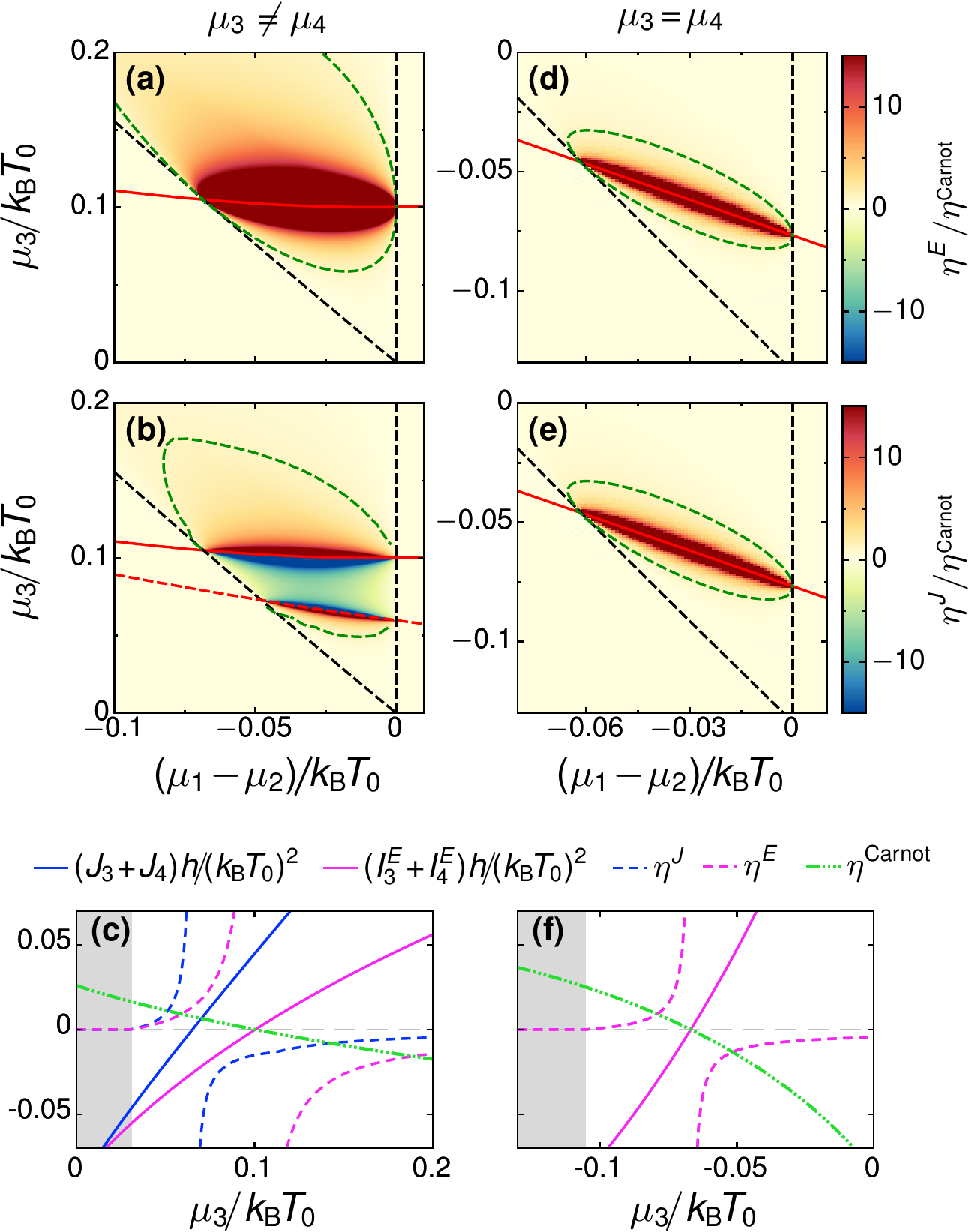}
\caption{Conventional efficiencies $\eta^E$ (see Eq.~(\ref{eq:effE})) and $\eta^J$ (see Eq.~(\ref{eq:effJ})) with respect to the Carnot bound of the setup producing power at $I_3+I_4=I_1+I_2=0$. A dashed green line shows the region within which the efficiencies exceed the Carnot limit. The solid red line indicates where the additional condition  $I^E_3+I^E_4=I_1^E+I_2^E=0$  is fulfilled. The dashed red line indicates where $J_3+J_4=0$.
For panels~(a) and (b) we fix $T_1=T_2=T_3= T_0$, but $T_4=1.2 T_3$ and take $\mu_4\neq\mu_3$ with $\mu_4/(k_{\text{B}}T_0)=-0.4$. 
For panels~(d) and (e) we  take $\mu_3=\mu_4$,  and temperatures $T_1=T_2=T_0$, $T_3/T_0=0.9$ and $T_4=1.2T_3$. For the line plots in panels~(c) and (f) we take the same parameter as in the columns above but fix $(\mu_1-\mu_2)/k_{\text{B}}T_0=-0.02$. The grey region indicates the region in which no power is produced.
\label{fig:standard_eff}}
\end{figure}

Comparing the produced power to the energy absorbed from the resource, yields the conventional efficiency defined in Eq.~(\ref{eq:effE}). The efficiency, compared to the Carnot limit, see Eq.~(\ref{eq:effCarnot}), is displayed in Fig.~\ref{fig:standard_eff}(a) and (d). Note that in order to define the Carnot bound for the conventional efficiency, a temperature has to be assigned to the distribution incoming from the resource region. This can be extracted from a probe (see Sec.~\ref{sec:probe}), leading to $\eta^\text{Carnot}=1-T_0/T_\text{p}$ with the probe temperature $T_\text{p}$.

The conventional efficiency $\eta^E$ is larger than the Carnot efficiency $\eta^\text{Carnot}$ in a large parameter regime (encircled by the dashed green line). This shows that the engine is not operated with heat as the \textit{only} resource, but that the nonequilibrium character of the incoming distribution is another resource.

What is more, the efficiency even diverges in some parameter intervals (on the solid red line between the two dashed black), because the sum of energy currents vanishes in the denominator.
However this is only part of the reason for the divergence seen on the solid red line in Figs.~\ref{fig:standard_eff}(a) and (d).
The second reason is that a probe will measure the temperature $T_\text{p}=T_0$ in this regime (because the probe must have the same distribution as terminal 2 if the distributions at the magenta and black circles in Fig.~\ref{fig:setup}(b) carry equal particle and energy currents). Thus the Carnot efficiency is zero on the solid red line.
This divergence of $\eta^E$ combined with the vanishing of $\eta^{\rm Carnot}$ can be seen in the line plots in panels (c) and (f), and leads to $\eta^E/\eta^{\text{Carnot}}\rightarrow\infty$. It means that our nonequilibrium device produces power even when no usual heat engine ---coupled to equilibrium baths at different temperatures--- would be able to do so.

Finally, we compare to a hypothetical efficiency defined via \textit{heat} currents as a resource, see Eq.~(\ref{eq:effJ}). The results for $\eta^J/\eta^{\text{Carnot}}$, as shown in Figs.~\ref{fig:standard_eff}(b) and (e), differ from $\eta^E/\eta^{\text{Carnot}}$ when there is an internal potential difference in the resource. Panel~(b) shows that the efficiency $\eta^J/\eta^\text{Carnot}$, can then both be positive and negative in regions, where power is \textit{produced} in the working substance. The reason for this is that the engine does useful work when heat is flowing either out of or into the resource region, but that heat and energy currents do not have the same sign. 
In the region of negative efficiencies $\eta^J/\eta^\text{Carnot}$ (blue-green part of the density plot), $\eta^J$ is negative, while $\eta^\text{Carnot}$ is positive. There the resource region is heated up due to Joule heating ($J_3+J_4>0$), while the energy current is flowing \textit{out of} it  ($I_3^E+I_4^E<0$).
The heat flow thus indicates that power production seemingly results from doing work with a cold bath. Instead, the flow of energy (and the resulting Carnot efficiency) indicate the opposite. 
These issues show that in a nonequilibrium device, it is a delicate task to identify meaningful resources 
 and the resulting efficiencies.

\subsection{Cooling}\label{sec:cooling}
 We now demonstrate that unexpected effects also appear when a nonequilibrium distribution is used to enable refrigeration. Here, we consider analogies to two types of refrigerators. This can be either a standard refrigerator exploiting electrical power to cool down a cold reservoir, or an absorption refrigerator, in which heat is used as a resource for cooling (cooling-by-heating). In order to be able to compare to usual machines, we therefore fix
 \begin{subequations}
  \begin{eqnarray}
  \text{standard\ refrigerator:}\ \  0&=& P_\text{in}-(J_1+J_2),\qquad \label{eq:cond_TEfridge}\\
     \text{absorption\ refrigerator:}\ \  0&=&I_3+I_4. \label{eq:cond_absfridge}
 \end{eqnarray}
\end{subequations}
The origins of these two equalities are explained above Eqs.~(\ref{eq:def_COP_TE}) and (\ref{eq:def_COP_abs}) respectively. The first law of thermodynamics means the first equality is equivalent to saying that there is no heat flow into (or out of) the nonequilibrium resource; $J_\text{neq}=0$.

\begin{figure}[t]
\includegraphics[width=\columnwidth]{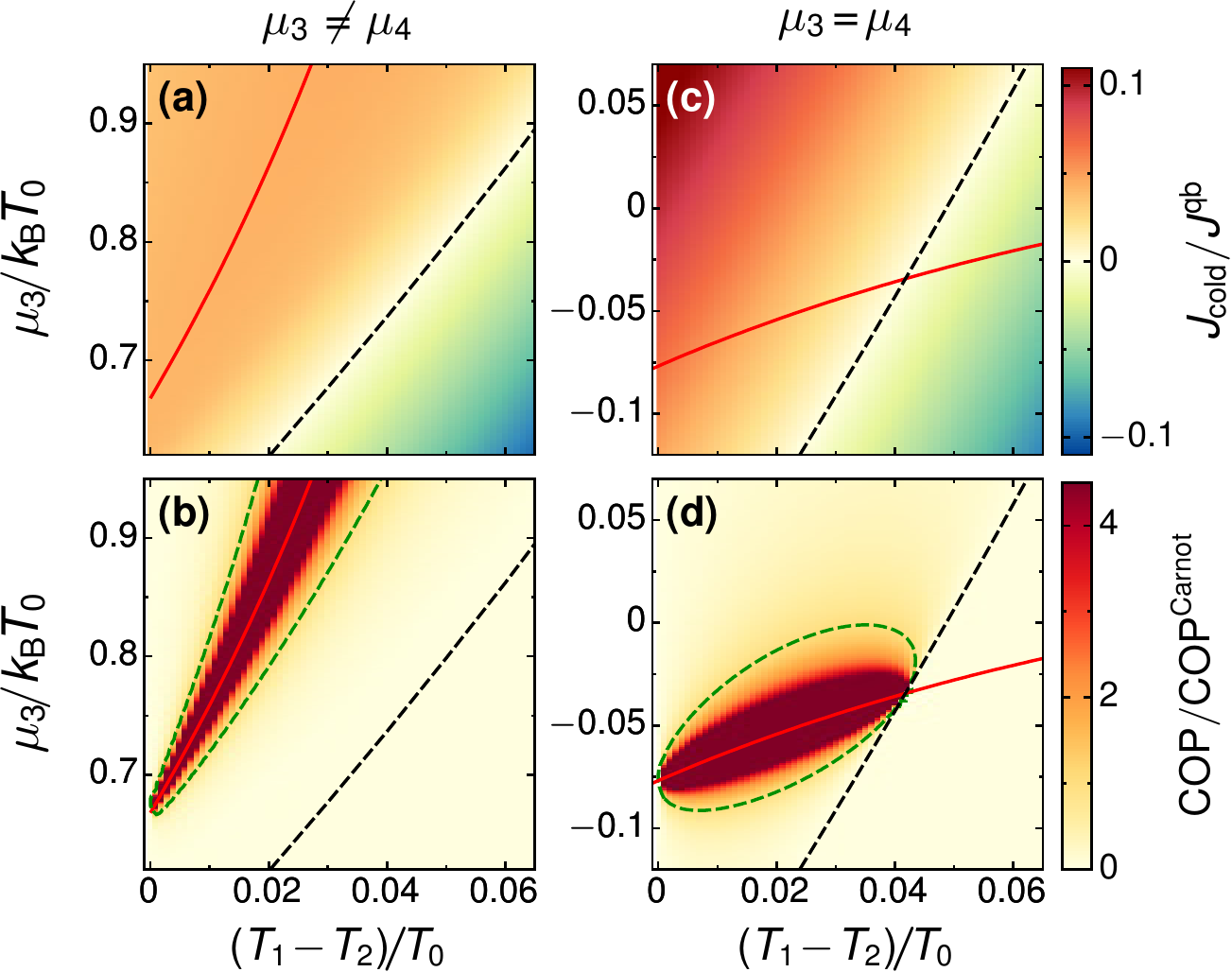}
\caption{Standard refrigerator: Cooling power, $J_\text{cold}=-J_2$ and conventional coefficient of performance with respect to its Carnot bound, see Eqs.~(\ref{eq:def_COP_TE}) and (\ref{eq:Carnot_COP_TE}), under the condition given in Eq.~(\ref{eq:cond_TEfridge}) and at $\mu_1=\mu_2=0$ and $T_1=T_0$. We fix $E_\text{ws}=0$ and adjust $E_\text{neq}$ such that Eq.~(\ref{eq:cond_TEfridge}) is fulfilled. Furthermore we fix in panels~(a) and (b)  $T_3=T_4=1.2T_0$ and $\mu_4/(k_{\text{B}}T_0)=-0.4$ and in panels~(c) and (d) $\mu_3=\mu_4$ as well as  $T_3=0.9T_0$ and $T_4=1.2T_0$.
The solid red line marks where the strict demon condition is fulfilled, $I^E_3+I^E_4=I_3+I_4=0$; the black dashed line shows where $J_{\rm cold}$=0; the dashed green line limits the region where COP$\geq$COP$^\text{Carnot}$. 
\label{fig:TEfridge}}
\end{figure}

We start by analyzing the analogue to a standard refrigerator, where we exploit the nonequilibrium distribution injected from the resource region to cool down reservoir 2, $J_\text{cold}=-J_2$.
Fig.~\ref{fig:TEfridge} shows the cooling power, as a function of the temperature bias between reservoir 1 and 2, and the electrochemical potential of reservoir 3.  The same principle could be used to cool terminal 1, instead; see the example of the absorption refrigerator below.
We consider both the case where the nonequilibrium distribution is created by a potential bias, $\mu_3\neq\mu_4$ (left column of Fig.~\ref{fig:TEfridge}), or by a temperature bias, 
$T_3\neq T_4$ (right column of Fig.~\ref{fig:TEfridge}), within the resource region. In both cases the system is able to cool down the cold reservoir, reservoir 2, in large parameter intervals, limited by the dashed line and the condition $T_1-T_2=0$. This is even possible in the absence of an overall charge current from the resource region, $I_3+I_4=0$, which we call a strict N-demon below (indicated by the solid red line).

The COP of the cooling process, see Eq.~(\ref{eq:def_COP_TE}), contains an electrochemical potential difference, which takes the form $\delta\mu=\mu_0-\mu_\text{p}$. The effective potential of the resource region, $\mu_\text{p}$, is extracted from the measurement of a voltage probe, as explained in Sec.~\ref{sec:probe}.

Panels~(b) and (d) of Fig.~\ref{fig:TEfridge} show that the COP can be larger than the Carnot bound to the conventional efficiency, given by  Eq.~(\ref{eq:def_COP_abs}). Also here, we even find cases of diverging COP, which is due to two reasons (similarly to the discussion of the divergences in the heat engine case). Along the solid red line, the COP diverges, because  the total charge current from the resource region is zero; at the same time, the electrochemical potential bias between working substance and resource region is zero at vanishing charge and energy flow, $\delta\mu=\mu_0-\mu_\text{p}=0$. In this case, the Carnot bound is independent of the effective temperature of the resource region and is therefore never suppressed to zero, as long as $T_\text{cold}=T_2\neq 0$.

\begin{figure}
\includegraphics[width=\columnwidth]{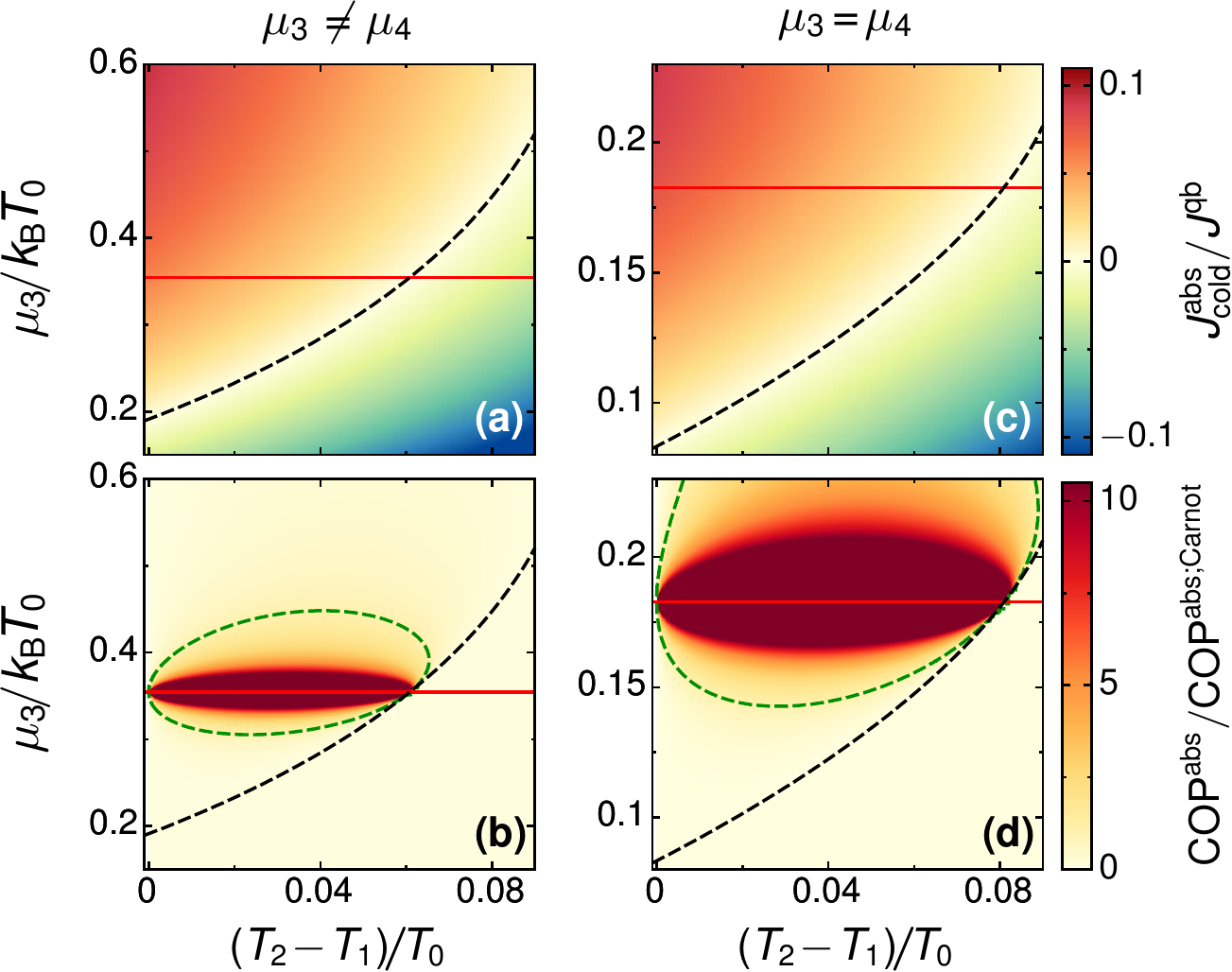}
\caption{Absorption refrigerator in the absence of charge currents from the resource, $I_3+I_4=I_1+I_2=0$, and for $\mu_1=\mu_2=0$ and $T_2=T_0$. We take $E_\text{ws}=0$ and fix $E_\text{neq}$ by the condition of vanishing charge currents.
We show the cooling power $J_\text{cold}^\text{abs}=-J_1$ and the conventional efficiency with respect to the Carnot bound as defined in Eqs.~(\ref{eq:def_COP_abs}) and (\ref{eq:Carnot_COP_abs}).
In panels~(a) and (b), we fix $T_3=T_4=1.2T_0$ as well as $\mu_4/(k_{\text{B}}T_0)=-0.8$. In panels~(c) and (d) we fix $T_3=1.15T_0$ and $T_4=0.8T_0$ as well as $\mu_4=\mu_3$.
At the solid red line the additional condition $I_3^E+I_4^E=I_1^E+I_2^E=0$ is fulfilled. The black dashed line indicates where $J_{\rm cold}$=0; the COP exceeds the Carnot bound within the dashed green limited region. 
\label{fig:absfridge}}
\end{figure}

In order to compare to an absorption refrigerator, we fix the total particle current between resource region and working substance to zero and use the nonequilibrium distribution to cool terminal 1. The result for the cooling power $J_\text{cold}^\text{abs}=-J_1$ is shown in Fig.~\ref{fig:absfridge}. It shows that also the analogue of an absorption refrigerator can be realized with our nonequilibrium device, leading to cooling at COPs above the conventional Carnot bound, see Eqs.~(\ref{eq:def_COP_abs}) and \eqref{eq:Carnot_COP_abs}. The divergence of the COP along the solid red line is again due to having $I_3^E+I_4^E\equiv0$, as well as a vanishing Carnot bound, since $T_\text{hot}\equiv T_\text{p}\rightarrow T_0$. Note that the red line, at which both charge and energy currents through the interface region are zero, is independent of the temperature bias in the working substance, since we have here chosen to fix $T_2\equiv T_0$.

\section{Two classes of demonic action}\label{sec:demonic_action}

The devices discussed in the previous sections exploit a nonequilibrium distribution to produce power or to cool a cold reservoir. As a result, the performance of the device as a heat engine or refrigerator seemingly violates thermodynamic bounds on efficiency; exceeding the Carnot efficiency and even diverging (often when the Carnot efficiency itself vanishes). These  \textit{N-demons}~\cite{Sanchez2019Nov} exploit nonequilibrium-ness as a resource, and so they function without or with a reduced inflow of power, heat or energy. Based on the results of Sec.~\ref{sec:noneq}, we now identify two different classes of such N-demons:
strict and relaxed.

\subsection{Strict N-demon}\label{sec:strict_demon}

We first discuss the \textit{strict} conditions, introduced for the N-demon in Ref.~\cite{Sanchez2019Nov}. Namely, the device functions even though the energy and  charge currents from the resource region to the working substance vanish:
\begin{align}\label{eq:strict_demon}
    \text{Strict N-demon:}\ \ & I_3+I_4=I^E_3+I^E_4\equiv0.
\end{align}
Part of the puzzle of this N-demon is resolved by noting that energy currents are conserved separately within the resource region and the working substance; hence the first law is not violated. Furthermore, entropy production in the resource region is different from zero even when heat currents vanish, thanks to its nonequilibrium character; hence the second law is not violated either.

The absence of charge and energy currents through the interface between resource region and working substance 
means that usual resources---namely energy and electrical power---are absent, while the produced power is still finite. Consequently, the efficiencies of the \textit{strict N-demon} working as a heat engine or refrigerator diverge,
\begin{align}
    \eta^E,\ \text{COP}^\text{},\ \text{COP}^\text{abs}& \rightarrow\infty.
\end{align}
Charge and energy currents vanish when the effective electrochemical potential and temperature of the resource region (as measured by the probe) equal $\mu_2$ and $T_2$. Interestingly, the divergence of efficiencies is accompanied by a vanishing Carnot bound for two devices,
\begin{align}
    \eta^\text{Carnot},\ \text{COP}^\text{abs;Carnot}& \rightarrow 0.
\end{align}
In contrast, the standard refrigerator's Carnot bound remains finite for being independent of the resource region.

Note that the strict N-demon, when operating as a heat engine, also maximally violates the quantum bound on power production in Eq.~(\ref{eq:WhitneyPower}). This bound equals zero when the effective temperature of the resource region equals $T_2= T_\text{cold}$, yet finite power production can occur, see Fig.~\ref{fig:power_production}. In contrast, the bound on cooling, see Eq.~(\ref{eq:pendrybound}) is never violated, since it is independent of the resource region. See also Figs.~\ref{fig:TEfridge} and \ref{fig:absfridge}, where the cooling power is normalized with respect to the bound.

\subsection{Relaxed N-demon}\label{sec:relaxed_demon}

The action of the multi-terminal device presented in this paper can be considered ``demonic" even when the strict demon conditions of Eq.~(\ref{eq:strict_demon}) are relaxed. 
Indeed, whenever the performance of the multi-terminal device exceeds the Carnot limit, the second law of thermodynamics might seem to be broken. The apparent violation is due to the nonequilibrium character of the resource. 

We define the conditions for a \textit{relaxed N-demon} as 
\begin{align}
\begin{array}{c}\text{Relaxed}\\ \text{N-demon:} \end{array} \ & 
\left\{ \begin{array}{cc}
{\displaystyle \frac{\eta^E}{\eta^\text{Carnot}} >1} & \ \hbox{heat engine}\\
\\
{\displaystyle \frac{\text{COP}^\text{}}{\text{COP}^\text{Carnot{}}}>1} &\ {\displaystyle 
\begin{array}{c}\text{standard}\\ \text{refrigerator} \end{array}}\\
\\
{\displaystyle \frac{\text{COP}^\text{abs}}{\text{COP}^\text{abs;Carnot}}>1} &\ {\displaystyle
\begin{array}{c}\text{absorption}\\ \text{refrigerator} \end{array}}
\end{array}\right.
\end{align}

The breaking of these usual thermodynamic bounds occurs for the following reasons. (i) The effective temperature and electrochemical potential (as measured by the probe) are not a true thermodynamic temperature and potential, because a nonequilibrium system has no true thermodynamic temperature and electrochemical potential. This leads to a breakdown of the Carnot expression for maximum efficiencies. (ii) In the conventional efficiencies, heat and electrical work figure as the resources for the engine operation. We have however seen that here the nonequilibrium state can also serve as a resource, with a nonvanishing entropy production.

In the following two sections, we will therefore introduce efficiencies which are more meaningful for the operation of the studied devices. We will furthermore show how thermalization of the resource suppresses the effect.

\section{Entropy coefficient}\label{sec:chi}

\begin{figure}[t]
\includegraphics[width=\columnwidth]{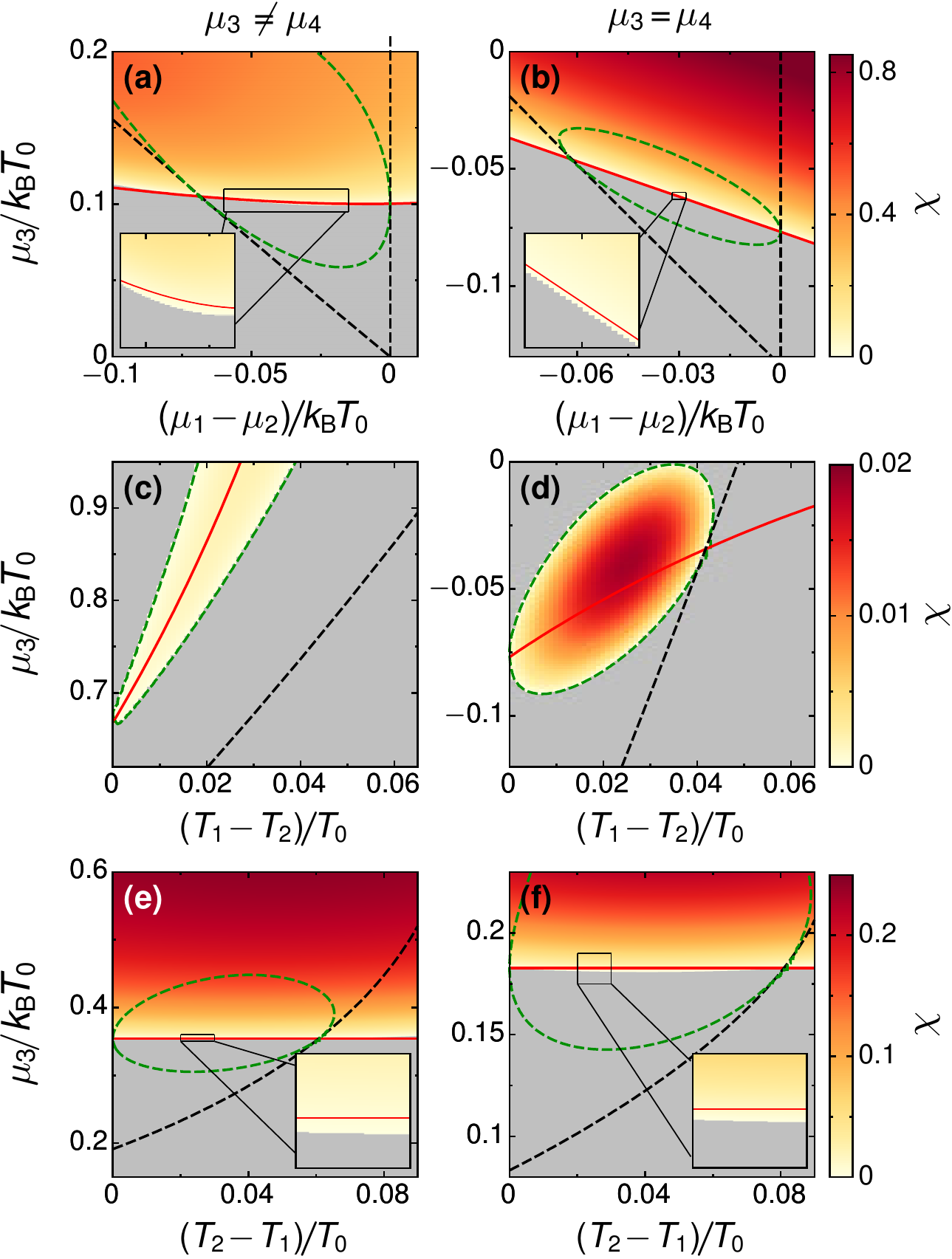}
\caption{Entropy coefficient $\chi$ for the heat engine and refrigerators, see Eq.~(\ref{eq:eff_entropy_2}). The solid red line indicates where the strict demon condition is fulfilled, $I^E_3+I^E_4=I_3+I_4=0$.
Panels~(a) and (b) show results for the analogue to a heat engine and all parameters are chosen as in Fig.~\ref{fig:power_production}. Panels~(c) and (d) show results for the standard refrigerator and all parameters are chosen as in Fig.~\ref{fig:TEfridge}. Panels~(e) and (f) show results for the absorption refrigerator and all parameters are chosen as in Fig.~\ref{fig:absfridge}.  Grey indicates that the entropy coefficient is negative.
\label{fig:ent_chi}}
\end{figure}

To capture the  entropy change associated with the nonequilibrium resource (which the conventional efficiencies overlook) we can define an entropy coefficient $\chi$.
It compares the entropy reduction in the working substance, $-\dot{S}_\text{ws}$ to the entropy production in the resource region, $\dot{S}_\text{neq}$, due to the action of the N-demon, as
\begin{equation}
\label{eq:eff_entropy_1}
    \chi=-\frac{\dot{S}_\text{ws}}{\dot{S}_\text{neq}}.
\end{equation}
This merits study because it  
reveals insight into the entropy balance in the device.  
However, it neglects the heat or work supplied by the non-equilibrium resource, so it is not a good measure of overall efficiency (for that, one should jump to Sec.~\ref{sec:free}).

The second law requires for any system that the total entropy production has to be positive, so~\cite{noheat}
\begin{align}
\label{eq:2ndlaw}
    \dot{S}_\text{neq}+\dot{S}_\text{ws}\geq0.
\end{align}
This sets a bound on the coefficient $\chi\leq1$. The coefficient $\chi$ is always positive when the nonequilibrium resource is consumed ($\dot{S}_\rmneq\geq0$), while power production or cooling of the cold terminal goes along with $\dot{S}_\rmws\leq0$.

Our simple choice for a nonequilibrium system, allows us to write simple expressions for $\dot{S}_\text{neq}$ and $\dot{S}_\text{ws}$. We use the Clausius relations to connect entropy production to heat flows, by using the fact that all terminals are at equilibrium.
We find 
\begin{equation}
\label{eq:eff_entropy_2}
    \chi=-\frac{J_1/T_1+J_2/T_2}{J_3/T_3+J_4/T_4}.
\end{equation}
Fig.~\ref{fig:ent_chi} plots this for the three types of machines. This clearly demonstrates that power production and cooling of the cold reservoir do not require a positive entropy coefficient. Only the strict N-demon always exhibits positive $\chi$  because it requires entropy production in the resource region and entropy reduction in the working substance, even though $\chi$ is often small  (see insets in Fig.~\ref{fig:ent_chi}). 

In Figs.~\ref{fig:ent_chi} (a), (b), (e) and (f), one can violate the Carnot bound with negative $\chi$. This corresponds to power production, while a small amount of entropy is generated in the working substance (negative numerator in $\chi$).
In contrast, for the analogue of the standard refrigerator in Fig.~\ref{fig:ent_chi}(c) and (d), the violation of the Carnot limit coincides with $\chi$ being positive. We can easily see that this is because the standard refrigerator conditions on heat currents impose that the numerator of Eq.~\eqref{eq:eff_entropy_2} vanishes at the points where $P_{\rm in}=(1-T_1/T_2)J_2$.

\section{Free-energy efficiency}\label{sec:free}

The conventional efficiencies bring out the unusual character of the nonequilibrium resource compared to well-known heat engines and refrigerators. At the same time, $\chi$  gives insight into  the entropy balance. However neither characterizes the device's true performance. For this we propose alternative efficiencies that
account for the nonequilibrium resource that is consumed by the device, in terms of changes in that \textit{nonequilibrium resource}'s free energy.

The coefficient $\chi$ contained the whole entropy production in the two device parts. However, to quantify the performance of steady-state power production, we are only interested in the part of the entropy deriving from the chemical work $\dot{S}^\text{power}=\mu_1 I_1/T_1\,+\,\mu_2 I_2/T_2$.
We consider the case where there is no temperature difference within the working substance, $T_1=T_2\equiv T_0$, so $\dot{S}^\text{power}=(\mu_1 I_1 +\mu_2 I_2)\big/T_0 = P/T_0$.
Assume furthermore, that the nonequilibrium resource supplies no particles, but supplies an energy $-(I^E_3+I^E_4)$. This is pure heat and  is gauge-invariant because~\cite{ingeneral}
$I_3+I_4=0$.
Then we can use the second law in Eq.~(\ref{eq:2ndlaw}), to write
\begin{eqnarray}\label{Eq:P-inequality}
P \leq  T_0\dot{S}_\text{neq} -(I^E_3+I^E_4)
\end{eqnarray}
Now we recall that the Helmholtz free energy (also called the Helmholtz potential, see e.g. sections 6-1 and 6-2 of Ref.~\onlinecite{Callen1985Sep}) of the nonequilibrium resource is~\cite{temperature}
$F_\text{neq}= U_\text{neq}-T_0 S_\text{neq}$, where $U_\text{neq}$ is the total energy of the nonequilibrium resource.
Noting that $\dot{U}_\text{neq}=(I^E_3+I^E_4)$,
we see that Eq.~(\ref{Eq:P-inequality}) reduces to
the statement that the upper bound on the power output is the rate of reduction of this free energy;
\begin{eqnarray}\label{Eq:P<free}
P &\leq&  -\dot{F}_\text{neq} 
\end{eqnarray}
As we are only  interested in this inequality when the power output $P>0$, we know the right-hand-side will be positive, so we can define a free-energy efficiency
\begin{eqnarray}\label{eq:eff_P_entropy}
\eta^\text{free}  &\equiv&  \frac{P}{-\dot{F}_\text{neq}} \ =\   \frac{P}{ T_0\dot{S}_\text{neq} -(I^E_3+I^E_4)} \ \leq\  1 .
\end{eqnarray}
For $P>0$, we can be sure that $0< \eta^\text{free} \leq 1$, giving it the usual properties of an efficiency.
Note that energy conservation means that one can replace $-(I^E_3+I^E_4)$ by $(I^E_1+I^E_2)$ in this formula, whenever that is more convenient.
Let us now examine this free-energy efficiency, $\eta^\text{free}$, in two limiting cases.
Firstly, consider the nonequilibrium resource is a strict N-demon (supplying no energy and no particles), then 
$\eta^\text{free} = P\big/(T_0\dot{S}_\text{neq})=\chi$, and   
the inequality tells us that $P \leq T_0\dot{S}_\text{neq}$, as given below Eq. (1) of Ref.~\cite{Sanchez2019Nov}.
Secondly, consider the nonequilibrium resource is  hot but \textit{not} out of equilibrium, so we have a conventional heat engine. For this  $T_3=T_4=T_\text{hot} > T_0$, and $\mu_3=\mu_4$ chosen to ensure no particle current through the interface. The resource gives energy $-(I^E_3+I^E_4)$ (this energy is heat, because the reservoir provides no work) to the working substance; then $\dot{S}_\text{neq}=(I^E_3+I^E_4)/T_\text{hot}$.  This means 
$\eta^\text{free} = P\big/[-(I^E_3+I^E_4) (1-T_0/T_\text{hot})]$, and the inequality in Eq.~(\ref{eq:eff_P_entropy}) gives the usual Carnot limit on a heat-engine's efficiency, cf. Eqs.~(\ref{eq:effE},\ref{eq:effCarnot}).

For the refrigerators, the aim is to reduce the entropy in the cold reservoir,
$\dot{S}^\text{cooling}=J_\text{cold}/T_\text{cold}$.
There is no bias between reservoirs 1 and 2 ($\mu_1{=}\mu_2{=}\mu_0$), and  
heat is removed from the colder of these two reservoirs at a rate $J_\text{cold}$, and (partially) pushed into the hotter reservoir of the working substance at rate $J_\text{hot}$.  
Remember our sign convention: if reservoir $l$ is being cooled, then we define $J_\text{cold}=-J_l$. 
The nonequilibrium resource drives a current $-(I_3{+}I_4)$ through the interface, thereby injecting an electrical power $P_\text{in}=-(\mu_\text{p}{-}\mu_1)(I_3{+}I_4)$ into the working substance.  
Defining the heat flow into the nonequilibrium resource as $J_\text{neq}=(I^E_3{+}I^E_4)-\mu_\text{p}(I_3{+}I_4)$, the first law reads 
$J_\text{hot}-J_\text{cold}+J_\text{neq}=P_\text{in}$.
The second law gives,
\begin{eqnarray}\label{Eq:Jcold-inequality}
J_{\rm cold}\left(\frac{T_\text{hot}}{T_\text{cold}} -1\right) \ \leq \ T_\text{hot}\dot{S}_\text{neq} -J_\text{neq}+P_\text{in} .
\end{eqnarray}
The Helmholtz free energy of the nonequilibrium system is~\cite{thot}
$F_\text{neq}= U_\text{neq}-T_\text{hot} S_\text{neq}$.
Noting that $\dot{U}_\text{neq}=J_\text{neq}-P_\text{in}$,
we see that Eq.~(\ref{Eq:Jcold-inequality}) reduces to
the statement that the upper bound on the cooling is given by the rate of reduction of this free energy;
\begin{eqnarray}
J_{\rm cold}\left(\frac{T_\text{hot}}{T_\text{cold}} -1\right) \ \leq \ -\dot{F}_\text{neq}.
\end{eqnarray}
As we are only interested in a situation in which the cold reservoir is being cooled,  $J_\text{cold}$ will be positive, and so the right-hand-side of the inequality must be positive as well.
Dividing through by the right-hand-side, enables us to define a free-energy coefficient of performance
\begin{eqnarray}\label{eq:COP_entropy_TE-general}
\text{COP}^\text{free} \,& \equiv &\,  \frac{J_{\rm cold}}{-\dot{F}_\text{neq}} 
\ =\ \frac{J_{\rm cold}}{T_\text{hot}\dot{S}_\text{neq} -J_\text{neq}+P_\text{in}}
\nonumber \\
&\leq&\, \frac{T_\text{cold}}{T_\text{hot}-T_\text{cold}}.
\qquad
\end{eqnarray}
At the same time $\text{COP}^\text{free}$ is never negative for cooling  (positive $J_\text{cold}$), giving it the usual properties of a COP.

An intriguing alternative formula for $\text{COP}^\text{free}$ can be found by eliminating $J_\text{neq}$ and $P_\text{in}$ using the first law, which reads $J_\text{hot}-J_\text{cold}+J_\text{neq}=P_\text{in}$.
Then
\begin{eqnarray}\label{eq:COP_entropy_TE-alternative}
\text{COP}^\text{free} 
\ =\ \frac{J_{\rm cold}}{T_\text{hot}\dot{S}_\text{neq} +J_\text{hot}-J_\text{cold}}.
\end{eqnarray}
This is harder to interpret than Eq.~(\ref{eq:COP_entropy_TE-general}), because it has  
the cooling power, $J_\text{cold}$, in both the numerator and denominator, however it will be useful in experimental situations in which $J_\text{neq}$ is hard to measure.

As for a textbook refrigerator, we take a situation where the nonequilibrium resource supplies electrical power $P_\text{in}$ but supplies no heat, $J_\text{neq}=0$.  However, it also supplies its nonequilibrium-ness as a resource to perform cooling.
We take $\mu_1=\mu_2$, so no work or cooling is done by the working substance in the absence of the nonequilibrium resource. Due to $J_\text{neq}=0$, the first law gives $J_\text{hot}=J_\text{cold}+P_\text{in}$, as in a textbook refrigerator, where $P_\text{in}=(\mu_\text{p}-\mu_0)(I_1+I_2)$, meaning that $I^E_1+I^E_2-\mu_\text{p}(I_1+I_2)$=0.
Then Eq.~(\ref{eq:COP_entropy_TE-general}) reduces to 
\begin{eqnarray}\label{eq:COP_entropy_TE-simpler}
\text{COP}^\text{free} \, = \, \frac{J_{\rm cold}}{T_\text{hot}\dot{S}_\text{neq}+P_\text{in}}
\ \leq \  \frac{T_\text{cold}}{T_\text{hot}-T_\text{cold}}.
\end{eqnarray}
Let us now examine two limiting cases.
Firstly, the nonequilibrium resource being a strict N-demon (supplying no energy and no particles), so the only thing it supplies is its nonequilibrium-ness. Then $P_\text{in}{=}0$, so
$\text{COP}^\text{free}{=}J_{\rm cold}\big/(T_\text{hot}\dot{S}_\text{neq})$, whose upper bound is $(T_\text{hot}/T_\text{cold}{-}1)^{-1}$.
Secondly, the resource is \textit{not} out of equilibrium ($T_3{=}T_4{=}T_\text{p}$), so $\dot{S}_\text{neq}{=}J_\text{neq}/T_\text{p}{=}0$ because $J_\text{neq}{=}0$. Then
$\text{COP}^\text{free} =-J_{\rm cold}/P_\text{in}$ coinciding with the usual COP; cf.\ Eqs.~(\ref{eq:def_COP_TE},\ref{eq:Carnot_COP_TE}).

Now we turn to absorption refrigerators, where the nonequilibrium resource supplies no electrical power $P_\text{in}=0$, because it supplies no particles $I_3+I_4=0$. Instead, it supplies heat $-J_\text{neq} = -(I^E_3+I^E_4)$. 
We consider cooling reservoir 1 (so $J_1=-J_\text{cold}$) moving the heat to reservoir 2 which is usually called ``ambient'' in such case (so $T_2=T_\text{amb}$).
The rate of change of free energy is $\dot{F}_\text{neq} = I^E_3+I^E_4 - T_\text{amb}\dot{S}_\text{neq}$.
Then Eq.~(\ref{eq:COP_entropy_TE-general}) gives

\begin{equation}
\text{COP}^\text{abs;free}{=} \frac{J_{\rm cold}}{T_\text{amb}\dot{S}_\text{neq} {-}(I^E_3{+}I^E_4)}\leq\frac{T_\text{cold}}{T_\text{amb}{-}T_\text{cold}},
\end{equation}
with $\text{COP}^\text{abs;free}$ always being positive for cooling  (positive $J_\text{cold}$).
In the limit of a strict N-demon, $\text{COP}^\text{abs;free}{=}-J_{\rm cold}\big/(T_\text{amb}\dot{S}_\text{neq})$.
In the opposite limit, where the resource is \textit{not} out of equilibrium ($\mu_3{=}\mu_4$ and $T_\text{p}{=}T_3{=}T_4$), but does supply heat ($T_\text{p}{>}T_\text{amb}$), we have 
$\dot{S}_\text{neq}= (I^E_3{+}I^E_4)/T_\text{p}$. Then 
$\text{COP}^\text{abs;free}=  J_{\rm cold}\big/\big[-(I^E_3+I^E_4) (1-T_\text{amb}/T_\text{p})\big]$ coincides with the usual COP for an absorption refrigerator, cf.\ Eqs.~(\ref{eq:def_COP_abs},\ref{eq:Carnot_COP_abs}).

\begin{figure}[t]
\includegraphics[width=\columnwidth]{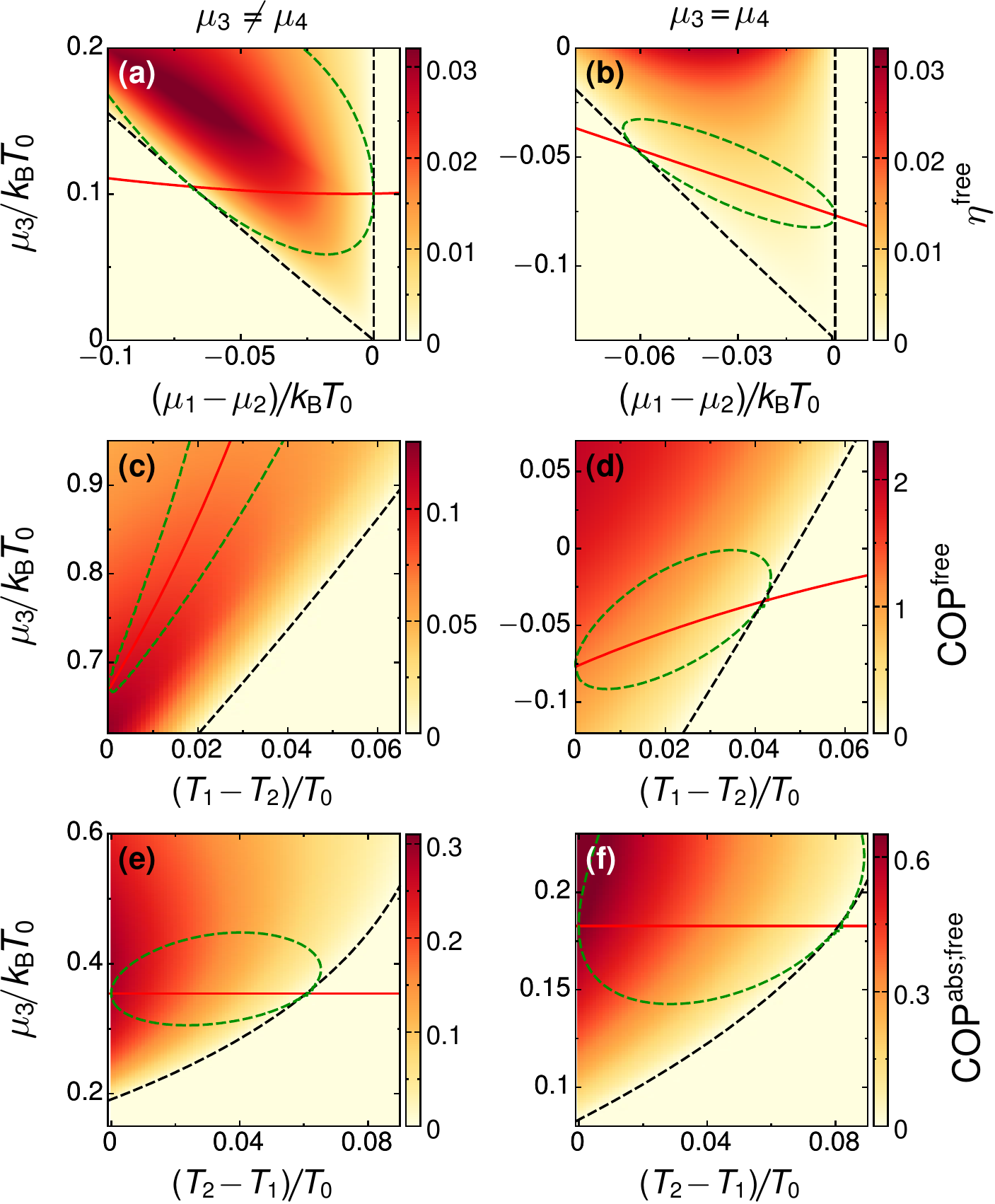}
\caption{Free-energy efficiencies for the heat engine $\eta^\text{free}$ and refrigerators $\text{COP}^\text{free}$ (see Eqs.~(\ref{eq:eff_P_entropy}) and (\ref{eq:COP_entropy_TE-simpler})). The solid red line indicates where the strict demon condition is fulfilled, $I^E_3+I^E_4=I_3+I_4=0$. 
For panels~(a) and (b) all parameters are chosen as in Fig.~\ref{fig:power_production}. For  panels~(c) and (d), all parameters are chosen as in Fig.~\ref{fig:TEfridge}. For panels~(e) and (f) all parameters are chosen as in Fig.~\ref{fig:absfridge}. 
\label{fig:ent_eff}}
\end{figure}

The results for these different free-energy efficiencies are shown in Fig.~\ref{fig:ent_eff}. The free-energy efficiencies are positive and can hence serve as useful efficiencies in the whole range of power production or cooling. 
Having shown that the free-energy efficiencies are useful quantities to characterize the performance of the multi-terminal device, we will in the following section propose it to optimize the device with respect to the choice of scattering probabilities $\tau_\text{neq}$ and $\tau_\text{ws}$.

In the future, it will be interesting to see if these free-energy efficiencies are useful for describing Otto engines that achieve large efficiencies by exploiting a non-equilibrium reservoir of squeezed-states~\cite{Niedenzu2018Jan,Ghosh2018Mar}.

\section{Thermalization}\label{sec:thermalize}

Here we show that (partial) thermalization of this nonequilibrium state deteriorates the device operation. 
We consider this to be an elegant proof that the effects that we discuss are due to the nonequilibrium-ness of the resource.  It also mimics the sort of experiment one would do to separate out the effects of nonequilibrium-ness on the operation of a device. 
One experimental way to do this would be to make the edge-state between the nonequilibrium resource and the working substance long enough that the nonequilibrium distribution thermalises before it gets to the working substance.  Another experimental possibility would be to 
use the voltage and temperature probe as introduced in Sec.~\ref{sec:probe}. In either case, we model the 
thermalisation using the voltage and temperature probe in Sec.~\ref{sec:probe}.

Consider a probe that completely thermalizes the distribution, $\tau_\text{p}=1$. 
We express $I^{(\nu)}_3+I^{(\nu)}_4$ in terms of the injected nonequilibrium function, as
\begin{align}
    I^{(\nu)}_3+I^{(\nu)}_4 & =\frac1h\int dE E^\nu\left(f_2(E)-\tilde{f}(E)\right)\nonumber\\
    & =\frac1h\int dE E^\nu\left(f_2(E)-f_\text{p}(E)\right), 
\end{align}
where in the second step, we have used the fact that the nonequilibrium distribution gets replaced by the probe distribution, $\tilde{f}(E)\rightarrow f_\text{p}(E)$, which carries the same amount of charge and energy. Importantly, the demon conditions $I^{(\nu)}_3+I^{(\nu)}_4 \equiv 0$ can then only be fulfilled for both energy and charge currents, if $f_2(E)=\tilde{f}(E)$.
This means that $I^{(\nu)}_2=-I^{(\nu)}_1$ always flows in the direction prescribed by the voltage and/or temperature difference in the working substance. 
Hence thermalization completely stops the strict N-demon from working; in this case the N-demon's only possible effect is to induce decoherence in the working substance.

\begin{figure}[t]
\includegraphics[width=\columnwidth]{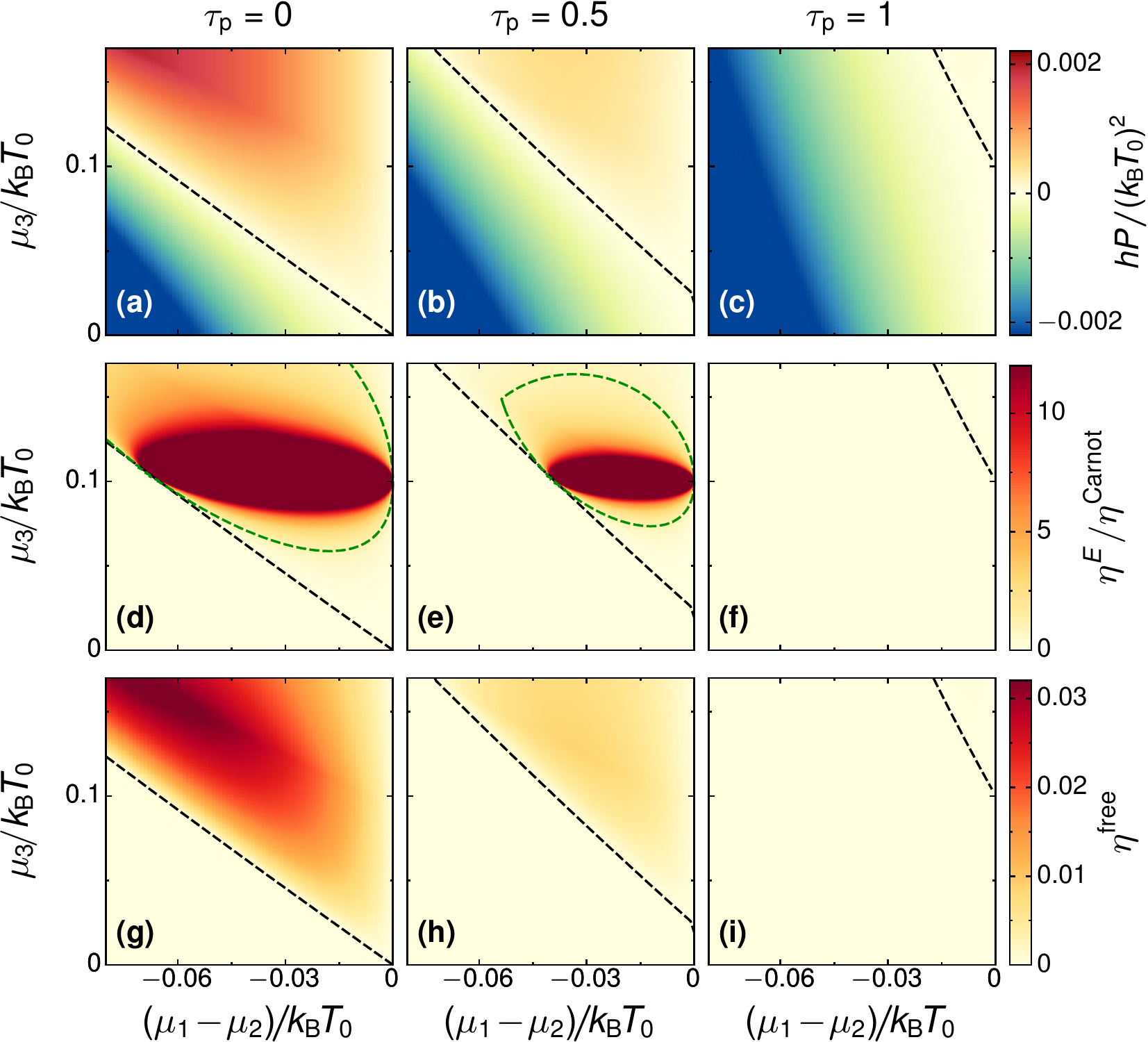}
\caption{Power production, conventional efficiency, and free-energy efficiency for the device working as a heat engine, at $I_3+I_4=I_1+I_2=0$. Here, a voltage and temperature probe is contacted to the conductor edge along which the nonequilibrium distribution is sent from the resource region to the working substance, with different coupling strengths $\tau_\text{p}$. All other  parameters are chosen as in Fig.~\ref{fig:power_production}(a). Black dashed lines indicate the (shrinking) region of finite power production, dashed green lines indicate the (shrinking) region of power production under relaxed demon conditions.
\label{fig:thermalize}}
\end{figure}

Fig.~\ref{fig:thermalize} shows that a partial thermalization by an incomplete coupling to the probe reduces the performance of the device. We focus on power production in analogy to a heat engine, showing the produced power, the conventional efficiency and the free-energy efficiency, for the example already shown
in Figs.~\ref{fig:power_production}(a) and \ref{fig:standard_eff}(a). With increasing thermalization (increasing coupling to the probe), the region in which power is produced shrinks, and the produced power is much smaller. At $\tau_\text{p}=1$, there is no power production under strict demon conditions, $I_3+I_4=I_3^E+I_4^E\equiv0$. The second row of Fig.~\ref{fig:thermalize} shows that the region in which the relaxed N-demon produces power shrinks with increasing thermalization. The conventional efficiency and free-energy efficiency are strongly suppressed, but remain finite in the small region of finite power production. This clearly shows that the partial thermalization of the incoming nonequilibrium distribution strongly suppresses both the strict and relaxed demonic effects. 
When the nonequilibrium resource is fully thermalized by the probe, the device behaves as a three-terminal thermoelectric device with the third terminal in internal equilibrium at temperature $T_\text{p}$~\cite{Sanchez2015Apr}.

\section{Exploring the parameter space of strict-N-demon devices}\label{sec:detailed_analysis}

After having shown the general operation of the four-terminal setup in the previous section, we now analyze in detail the performance of the device acting as a ``strict N-demon". We focus in particular on an analysis of how the properties of the three scattering regions, $\tau_\text{ws},\tau_\text{neq}$ and $\tau_\text{int}$ influence the demonic action.

\subsection{Constraints on transmission probabilities in the linear-response regime}\label{sec:tau_constraints}

In the  linear-response regime, chemical potential and temperature differences are small compared to $\kBT$. We focus on the linear-response regime for two reasons. Firstly, it is possible to achieve analytical results that shed light on the role of the three scatterers. Secondly, in this regime, one might expect close similarities to a usual thermoelectric device, since the nonequilibrium distribution in the resource region only weakly deviates from a thermal distribution. The response of conventional three-terminal thermoelectrics relies on the breaking of electron-hole symmetry by energy-dependent scatterers; hence we start by checking if this is required in our four-terminal setup under strict demon conditions.

In the linear-response regime, one can expand the charge and heat currents 
in terms of the electric and thermal affinities $\mathcal{A}_\alpha^\mu{=}\mu_\alpha/\kBT_0$ with respect to a reference potential $\mu_0\equiv0$ and $\mathcal{A}_\alpha^T{=}(T_\alpha-T_0)/\kBT_0^2$.
It is useful to define the following integrals:
\begin{subequations}
\begin{align}
\label{eq:gin}
g_{i}^{(\nu)}&=\int dEE^\nu{\tau}_{i}(E)\xi(E),\\
\label{eq:hin}
h_{i}^{(\nu)}&=\int dEE^\nu{\tau}_{\rm int}(E){\tau}_i(E)\xi(E),\\
\label{eq:qin}
q^{(\nu)}&=\int dEE^\nu{\tau}_{\rm ws}(E){\tau}_{\rm int}(E){\tau}_{\rm neq}(E)\xi(E),
\end{align}
\end{subequations}
where $\nu=0,1,2$ and $i\in\{\text{int,neq,ws}\}$. Here, we have defined $\xi(E){=}{-}\kBT_0 df_0(E)/dE$, with the Fermi function $f_0(E)$ with temperature $T_0$ and Fermi energy $\mu_0=0$.
The coefficients $g_i^{(1)}$ of each scattering region give the Seebeck and Peltier coefficients of its linear thermoelectric response~\cite{Sivan1986Jan,Butcher1990Jun}.
With these, we write the expressions for all particle and energy currents. In the resource region,
\begin{subequations}
\begin{align}
I_3^{(\nu)}&=\frac{1}{h}\left[g_{\rm int}^{(\nu)}\mathcal{A}_{23}^\mu+\left(g_{\rm neq}^{(\nu)}-h_{\rm neq}^{(\nu)}\right)\mathcal{A}_{43}^\mu\right.\\
&\left.+g_{\rm int}^{(\nu{+}1)}\mathcal{A}_{23}^T+\left(g_{\rm neq}^{(\nu{+}1)}{-}h_{\rm neq}^{(\nu{+}1)}\right)\mathcal{A}_{43}^T\right]\nonumber\\
I_4^{(\nu)}&=-\frac{1}{h}\left[g_{\rm neq}^{(\nu)}\mathcal{A}_{43}^\mu+g_{\rm neq}^{(\nu{+}1)}\mathcal{A}_{43}^T\right],
\end{align}
\end{subequations}
where $\mathcal{A}_{\alpha\beta}^\mu=\mathcal{A}_\alpha^\mu-\mathcal{A}_\beta^\mu$ and $\mathcal{A}_{\alpha\beta}^T=\mathcal{A}_\alpha^T-\mathcal{A}_\beta^T$. 
Importantly, the coefficients $g_i^{(1)}=0$, $h_i^{(1)}=0$ and $q^{(1)}=0$ vanish, if the transmission probabilities in the respective integrals are energy-independent or if they or their product are electron-hole symmetric. 
In particular, this means that if both scattering regions, with $\tau_{\rm int}$ and $\tau_{\rm neq}$, are either energy-independent or electron-hole symmetric, the particle currents become decoupled from the temperature bias and the energy currents become decoupled from the potential bias. Hence, the strict demon conditions, Eq.~\eqref{eq:strict_demon}, for electron-hole symmetric $\tau_{\rm int}$ and   $\tau_{\rm neq}$ are 
\begin{subequations}\label{eq:simple_linear_demon}
\begin{align}
    g^{(0)}_\text{int}\mathcal{A}_{23}^\mu+h^{(0)}_\text{neq}\mathcal{A}_{34}^\mu&\equiv0\\
    g^{(2)}_\text{int}\mathcal{A}_{23}^T+h^{(2)}_\text{neq}\mathcal{A}_{34}^T&\equiv0
\end{align}
\end{subequations}
which come respectively from the charge and energy currents out of the demon region. This shows that when the electrochemical potentials are tuned to fulfill the demon conditions for given $T_3$ and $T_4$, then either $\tau_{\rm int}$ or $\tau_{\rm neq}$ must break electron-hole symmetry. Otherwise, as Eqs.~(\ref{eq:simple_linear_demon}) show, the demon conditions can not be adjusted in linear response by uniquely tuning the electrochemical potentials $\mu_3$ and $\mu_4$. See a discussion of the requirements for this specific case in Appendix~\ref{app:linear}. For the rest of this section, we only require that $h^{(0)}_\text{neq}$ and $h^{(2)}_\text{neq}$ are finite in order to obtain sufficient tunability of the current flows to fulfill the strict demon conditions.

We now focus on a strict N-demon producing power and therefore assume $T_1{=}T_2$. Then $I_1{=}{-}I_2$, and we find
\beq
I_2=\frac{1}{h}g_{\rm ws}^{(0)}\mathcal{A}_{12}^\mu+I_{2,{\rm nonloc}},
\eeq
divides into the usual contribution and a term, nonlocally induced by the nonequilibrium distribution from the resource region. For electron-hole symmetric transmissions $\tau_\text{int}$ and $\tau_\text{neq}$, the nonlocal part reads:
\beq
I_{2,{\rm nonloc}}=\frac{1}{h}\left(C_{00}\mathcal{A}_{32}^\mu+C_{12}\mathcal{A}_{3}^T\right),
\eeq
with 
$C_{\nu\nu'}=\big(q^{(\nu)}g_{\rm int}^{(\nu')}-h_\rmws^{(\nu)}h_\rmneq^{(\nu')}\big)\big/h_\rmneq^{(\nu')}$.
This means that $I_{2,{\rm nonloc}}=0$ if either $\tau_{\rm int}$ and $\tau_\rmneq$, or $\tau_{\rm int}$ and $\tau_\rmws$ are simultaneously \textit{energy-independent}. However, there is no requirement for electron-hole symmetry breaking, in contrast with a usual thermoelectric device.
For the special case of $\tau_\text{int}\equiv1$, this has as an implication that both $\tau_\text{ws}$ and $\tau_\text{neq}$ need to be energy-dependent in the linear-response regime.

In nonlinear response, these constraints are partially lifted as seen in Secs.~\ref{sec:nonlinear_constraints} and \ref{sec:parameters}.

\subsection{Minimal energy-dependence in the nonlinear regime}\label{sec:nonlinear_constraints}

In the nonlinear regime, for any type of device, the demonic effect also requires a minimal energy-dependence of the transmission probabilities. 
The strict demon conditions, $I_3^{(\nu)}+I_4^{(\nu)}=0$,  lead to
\begin{align}
    &\hskip -3mm\int dE E^\nu\tau_\text{int}(E)\left[f_2(E)-f_3(E)\right]\nonumber\\
    &= \int dE E^\nu\tau_\text{neq}(E)\tau_\text{int}(E)\left[f_4(E)-f_3(E)\right].\label{eq:explicit_demon}
\end{align}
This means that the strict demon conditions can be non-trivially fulfilled, even if the transmission probabilities $\tau_\text{neq}, \tau_\text{int}$ are energy-independent.
However, if the transmission probability of the working substance is energy-independent ---for arbitrary transmission probabilities $\tau_\text{neq}, \tau_\text{int}$--- we find the following. Plugging the expressions obtained from the strict demon condition, Eq.~(\ref{eq:explicit_demon}), and for $\tilde{f}(E)$, Eq.~(\ref{eq:ftilde}), into Eq.~(\ref{eq_I1}), gives
\begin{align}
    I_1^{(\nu)} &= \frac{\tau_\text{ws}}{h}\int dE E^\nu\left[f_2(E)-f_1(E)\right]\ .
\end{align}
These charge and energy currents always flow into the direction imposed by the voltage and temperature bias in the working substance. This means that to perform work with a nonequilibrium distribution under strict demon conditions, one requires the transmission probability of the scatterer in the working substance to be energy-dependent. However, it is not required that this breaks the electron-hole symmetry of the transmission!

\subsection{Different device realizations}\label{sec:parameters}

Now we turn to specific implementations of the scatterers using QPCs or quantum dots.  We focus on the device producing electrical power under strict demon conditions.   We consider the most easily accessible experimental regime, in which all reservoirs except one are at the same temperature, $T_1=T_2=T_3\neq T_4$, as in Figs.~\ref{fig:power_production}(a), \ref{fig:standard_eff}(a), and \ref{fig:ent_eff}(a). The 
bias in the resource region $\mu_3\neq \mu_4$ then ensures the strict demon conditions.

We analyze the produced power, as well as the free-energy efficiency defined in Eq.~(\ref{eq:eff_P_entropy}). 
We note that under strict demon conditions and for $T_1=T_2$, so $\eta^\text{free}$ equals the entropy coefficient $\chi$.

\subsubsection{Both scatterers defined by QPCs}\label{sec:QPCQPC}

We start with both scatterers, that in the resource region and that in the working substance, being QPCs. The analysis in Sec.~\ref{sec:noneq} fixed $E_\text{ws}=0$ and adapted $E_\text{neq}$ in order to fulfill the heat engine condition of zero energy currents between resource region and working substance. Here, instead, we adapt $\mu_3$ and $\mu_4$ to fulfill the strict demon conditions and analyze the device's performance as a function of the scatterer characteristics.

\begin{figure}[tb]
\includegraphics[width=\columnwidth]{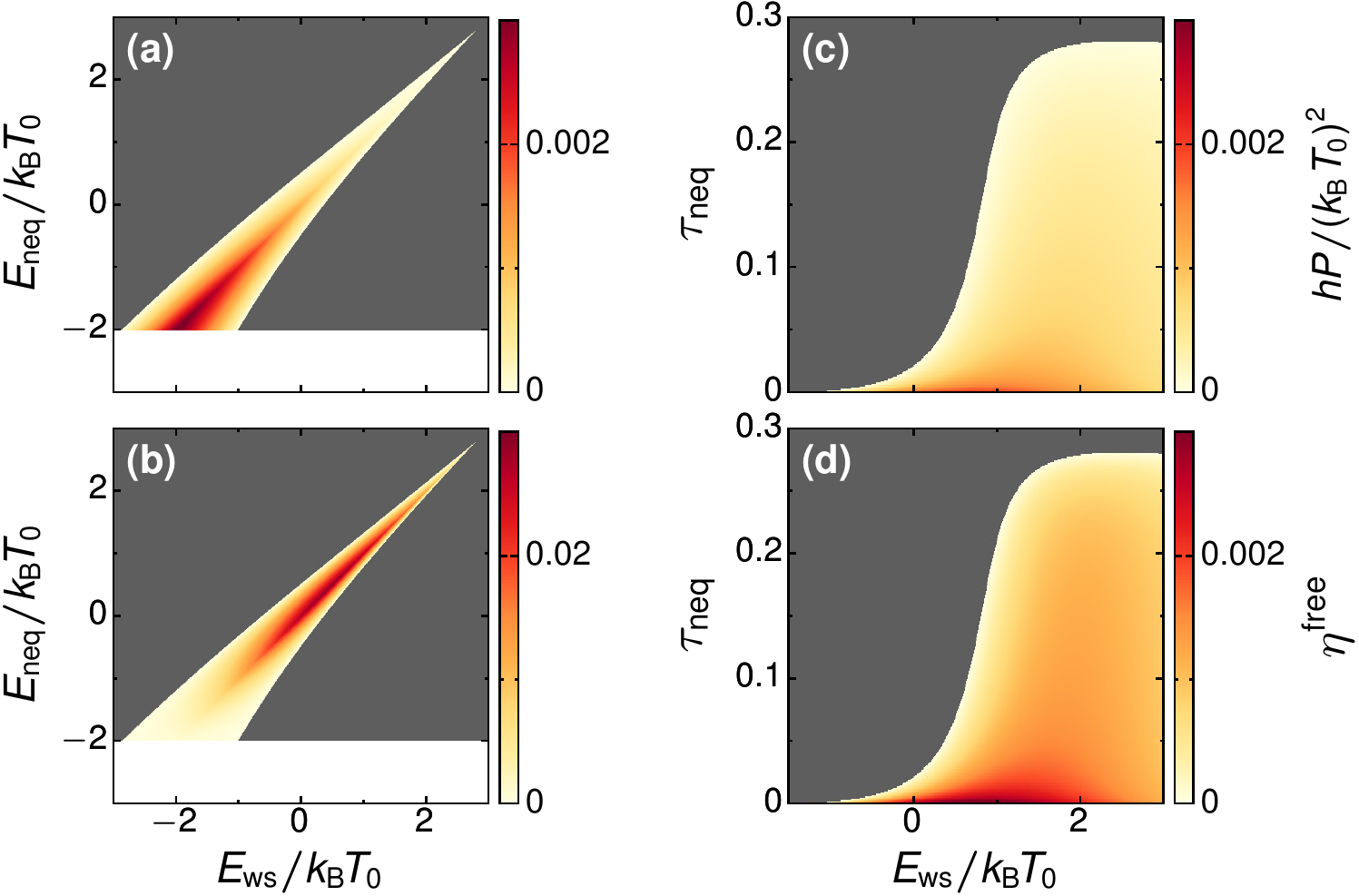}
\caption{Power production under strict demon conditions when both scatterers are QPCs. In the grey region no power is produced; in the white region the demon conditions can not be satisfied. (a), (b) Power production and free-energy efficiency for $\tau_\text{neq}(E)=\tau_\text{QPC}(E,E_\text{neq})$ and $\tau_\text{ws}(E)=\tau_{\rm QPC}(E,E_\text{ws})$; we fix  $T_1=T_2=T_3=T_0$. The same quantities are shown in (c) and (d) for $\tau_\text{neq}(E)\rightarrow\tau_\text{neq}$ and $\tau_\text{ws}=\tau_{\rm QPC}(E,E_\text{ws})$; we fix  $T_1=T_2=T_0$, $T_3=0.9T_0$ and $T_4=1.2T_0$. 
For all plots $(\mu_1-\mu_2)/(k_{\text{B}}T_0)=-0.05$  to ensure large  power production, and $\mu_3\neq\mu_4$ are fixed by the strict demon conditions.
\label{fig:QPCQPC}}
\end{figure}

The results are shown in Fig.~\ref{fig:QPCQPC}(a) and (b) for the QPC transmissions assumed to be described by sharp steps, cf. Eq.~\eqref{eq:qpc}. The power production is optimal when both characteristic energies are chosen to be approximately equal, $E_\text{neq}\simeq E_\text{ws}$. This optimal case is similar to a trivial realization of the demon situation, where the part of the distribution injected from terminal 3 ends up in terminal 2, while the part of the nonequilibrium distribution injected from terminal 4 ends up in 1. However, this is not a requirement for the device to function. The power production is largest when the step energies are low with respect to the mean potential in the working substance (as long as the strict demon conditions can be fulfilled).
Similar to what is expected from usual heat engines, the efficiency gets larger with suppressed power production. The efficiency also has its maximum in the vicinity of $E_\text{neq}\simeq E_\text{ws}$, but for $E_\text{neq},E_\text{ws}\geq0$. 

The device still functions when the transmission of the QPC in the resource region is so smooth that its energy dependence can be fully neglected, see Figs.~\ref{fig:QPCQPC}(c) and (d). In this case the intuitive picture of dividing a distribution into parts does not hold any longer, as mentioned above. Interestingly, the regimes of large power production and free-energy efficiency do then coincide at very weak transmissions $\tau_\text{neq}\ll1$.
Note that the maximum efficiencies that can be reached here are by an order of magnitude smaller compared to the QPC with a sharp transmission step in the resource region. In other words, a larger amount of entropy is produced in the resource region if the mixing is not energy-selective. 

Overall, the plots in Fig.~\ref{fig:QPCQPC} show that both the power production and the free-energy efficiencies are generally rather low under strict demon conditions. The main advantage from a device point of view rather lies in the fact that power production goes along with an entropy \textit{reduction} in the working substance. The entropy production, localized in the resource region, is well separated from the part of the device where the work is done.

\subsubsection{Both scatterers defined by quantum dots}\label{sec:QDQD}

\begin{figure}[tb]
\includegraphics[width=\columnwidth]{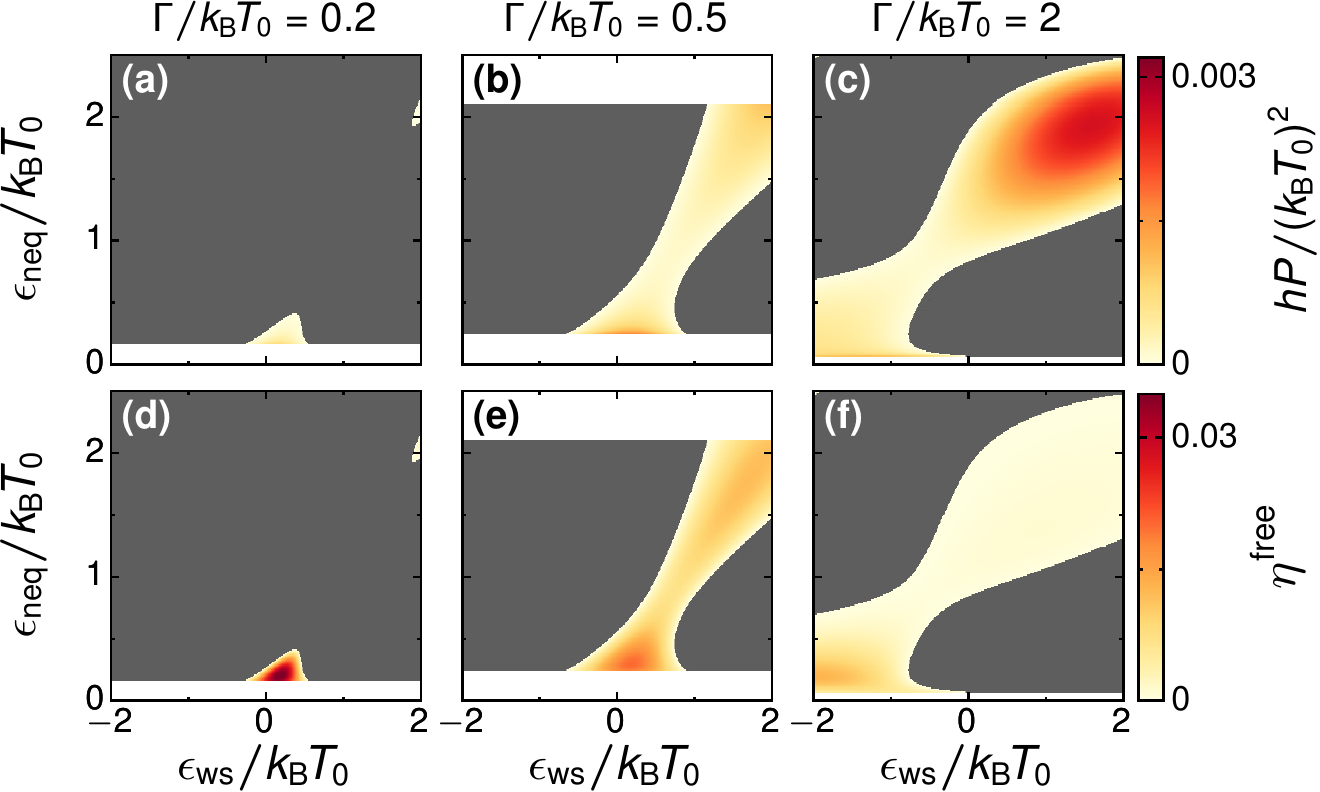}
\caption{(a)-(c) Power production and (d)-(f) free-energy efficiency under strict demon conditions when both scatterers are quantum dots with $\tau_\text{neq}(E)=\tau_{\rm qd}(E,\epsilon_\text{neq},\Gamma)$ and $\tau_\text{ws}(E)=\tau_{\rm qd}(E,\epsilon_\text{ws},\Gamma)$, for different $\Gamma$s. In the grey region no power is produced; in the white region the demon conditions can not be satisfied. 
We fix $T_1=T_2=T_3=T_0$, $T_4=1.2T_0$, and tune $(\mu_1-\mu_2)/(k_{\text{B}}T_0)=-0.05$ to ensure large power production, with $\mu_3\neq\mu_4$ fixed by the strict demon conditions.
\label{fig:QDQD}}
\end{figure}

Here we consider that both scatterers are quantum dots with Lorentzian-shaped transmissions, as given by Eq.~(\ref{eq:qd}).  To limit the parameter space's size, we take both quantum dots to have tunnel coupling $\Gamma$, and 
Fig.~\ref{fig:QDQD} show the effect of varying $\Gamma$. 
The strict demon conditions restrict power production to specific parameter regions, and the plots show how intricately the shape of these regions depends on the  parameters of the scatterers.

Similar to quantum dot thermoelectrics~\cite{Nakpathomkun2010Dec}, the efficiency is largest when the transmission probability is given by a sharp resonance. In contrast, the power production increases with increasing broadening of the resonance. The regions of power production under strict demon conditions are very limited for sharp resonances and become more extended when the resonances broaden.

There is however one important difference from usual thermoelectric devices: here, power production is possible, even if the level position of the dot in the working substance is at the electron-hole symmetric point, $\epsilon_\text{ws}=0$, see also the discussion in Secs.~\ref{sec:tau_constraints} and \ref{sec:nonlinear_constraints}.

In Fig.~\ref{fig:QDQD}, we have $\mu_1-\mu_2=-0.05\kBT_0$. However, the power production is the same if one simultaneously changes the sign of  $\epsilon_\text{ws}$, $\epsilon_\text{neq}$ and $\mu_1-\mu_2$.

\subsubsection{Device with quantum dot and QPC}\label{sec:QDQPC}

Obviously, the type of scatterer used in the resource region for mixing the nonequilibrium does not have to be the same as the one used in the working substance. In possible device applications, one might even simply get a nonequilibrium distribution as a ``waste" from some other operation. In this case only the working substance can be designed to optimize the power production. 

Here we take one of the scatterers to be a QPC with a sharp step-like transmission while the other one is a quantum dot with a finite broadening $\Gamma=0.5k_\text{B}T_0$. The goal is to determine which one is best for the resource region and which is best for the working substance.

\begin{figure}
\includegraphics[width=\columnwidth]{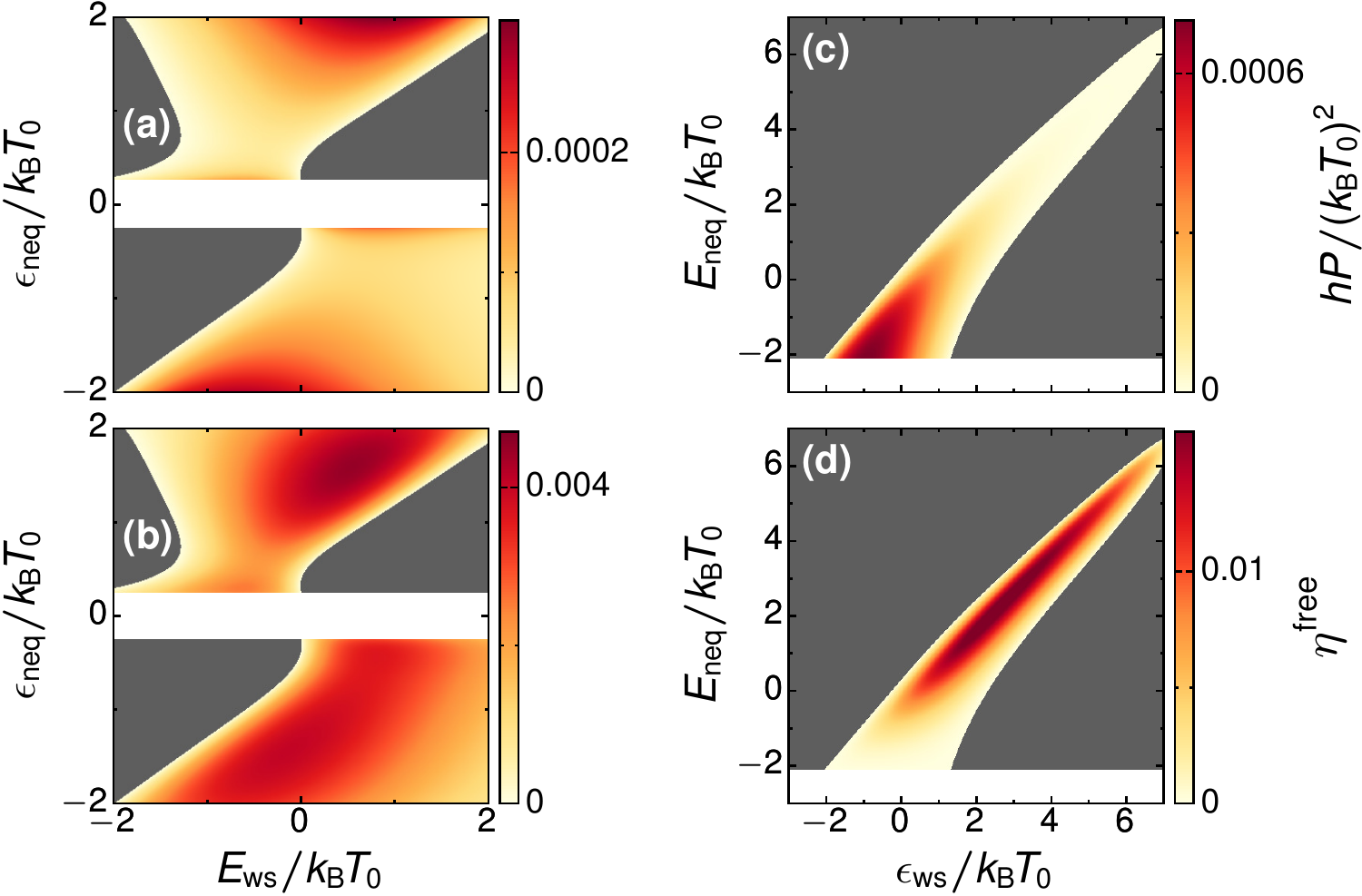}
\caption{Power production and free-energy efficiency under strict demon conditions where scatterers are a quantum dot or a QPC. (a), (b): $\tau_\text{neq}(E)=\tau_{\rm qd}(E,\epsilon_\text{neq},\Gamma)$ and $\tau_\text{ws}(E)=\tau_{\rm QPC}(E,E_\text{ws})$. (c), (d): $\tau_\text{neq}(E)=\tau_{\rm QPC}(E,E_\text{neq})$ and $\tau_\text{ws}(E)=\tau_{\rm qd}(E,\epsilon_\text{ws},\Gamma)$. In the grey region, no power is produced; in the white region the strict demon conditions can not be satisfied. We fix $T_1=T_2=T_3=T_0$, $T_4=1.2T_0$ and $\Gamma/(k_{\text{B}}T_0)=0.5$. We tune $(\mu_1-\mu_2)/(k_{\text{B}}T_0)=-0.01$ to ensure large power production, with $\mu_3\neq\mu_4$ fixed by the strict demon conditions.
\label{fig:QPCQD}}
\end{figure}

Fig.~\ref{fig:QPCQD} (a) and (b) show the case with a quantum dot in the resource region, and a sharp-step QPC in the working substance. If the level of the quantum dot is above the average electrochemical potential of the working substance, then power production is possible only if the step in the QPC transmission lies above $\epsilon_\text{neq}$. The opposite applies for negative characteristic energies. Both the power production and efficiency are very small, however the regions of maximum power production and maximum efficiencies can now be made to overlap in large regions.

Fig.~\ref{fig:QPCQD} (c) and (d) show the results for a QPC placed in the resource region, and a quantum dot in the working substance. In this case, the dot level $\epsilon_\text{ws}$ always needs to be below the step energy of the QPC, $E_\text{neq}$. The situation looks otherwise rather similar to the case of QPCs only, where large power and large efficiencies are realized in opposing parameter regimes close to $\epsilon_\text{ws}\simeq E_\text{neq}$. While the power production is of similar size compared to the ``swapped" case, the efficiency is increased by more than a factor of 2 compared to panel~(b).

\section{Conclusions}\label{sec:conclusions}

We have analyzed a multi-terminal quantum Hall setup, which exploits the nonequilibrium distribution created in one part of the device (the resource region) to produce power or to cool a reservoir in another part of the device (the working substance). We go beyond the N-demon proposed in our Ref.~\cite{Sanchez2019Nov}, which operated in the absence of charge and energy currents between the resource region and the working substance (we now call this a ``strict N-demon"). Here, we consider situations in which the resource and working substance can exchange energy or particles, calling this a ``relaxed N-demon''.  There we find the nonequilibrium character of the resource induces conventional efficiencies beyond the Carnot limit. The reason for this is that the nonequilibrium-ness (neglected in conventional definitions of efficiency) should be considered as a resource. Thus here we propose and investigate ``free-energy efficiencies'', based on the idea that a free energy change for the nonequilibrium resource is a comprehensive method of characterizing \textit{all} resources consumed (heat, power and nonequilibrium-ness).

We have studied this physics theoretically in a four-terminal set-up, for a broad range of parameters. 
We have shown that adding processes that destroy the nonequilibrium character of the injected distribution (such as thermalization by a voltage and temperature probe) destroys the demonic effects. 
The free-energy efficiencies have allowed us to analyze the performance of the device in a broad range of parameter regimes
that are relevant for future experiments in quantum Hall devices.  At the same time, we anticipate that our free-energy efficiencies could equally describe other nonequilibrium electronic and optical devices.

Over recent years, works on 
three-terminal systems (introduced in Refs.~\cite{Sanchez2011Feb,Entin-Wohlman2012Feb,Jiang2012Feb} and explored experimentally in Refs.~\cite{Roche2015Apr,Hartmann2015Apr,Thierschmann2015Aug}) 
have addressed definitions of efficiencies.
The correct choice of definition is difficult in situations where there are either multiple inputs  
(e.g. multiple heat sources) \cite{Mazza2014Aug,Jiang2014Nov}
or multiple outputs 
(a  combination of power production and cooling \cite{Entin-Wohlman2015Feb}, multiple power outputs \cite{Jiang2014Oct,Lu2017Jul,Lu2019Sep}), or more general hybrid thermal machines~\cite{manzano2020}.
We suspect that the free energy efficiency used in the present work could be adapted to be a good measure of efficiencies
in such situations as well.

{\it Note added---} While finishing this manuscript we became aware of a work~\cite{Deghi2020Jul} which addressed a similar model, and introduced efficiencies like in our Sec.~\ref{sec:free}. Shortly after this work was submitted, Ref.~\onlinecite{Jiang2020Aug} appeared discussing the same physics, but in a four-terminal phonon-electron system.

\acknowledgements

We thank Juliette Monsel and Jens Schulenborg for useful comments on the manuscript. This research was supported in part by the National Science Foundation under Grant No. NSF PHY-1748958 (RS and JS), the Knut and Alice Wallenberg Foundation and the Vetenskapsr\r{a}det, Swedish VR (JS and FH), the Ram\'on y Cajal program RYC-2016-20778, the 
Spanish Ministerio de Ciencia e Innovaci\'on via grant No. PID2019-110125GB-I00 and through the ``Mar\'ia de Maeztu'' Programme for Units of Excellence in R{\&}D CEX2018-000805-M (RS), and the French Agence Nationale de la Recherche under the programme ``Investissements d'avenir'' ANR-15-IDEX-0002 (RW).

\appendix

\section{Useful integrals}
\label{app:integrals}

In the case where the scatterer is realized by a QPC and the transmission coefficients are given by a sharp step functions, $\tau_\text{QPC}=\Theta(E-E_0)$, the charge and heat currents are linear combinations of integrals of the form 
\beq
{\cal I}_{\alpha,\beta}^{(\nu)}(E_0)=\int_{E_0}^\infty dEE^\nu(f_\alpha(E)-f_\beta(E)).\label{eq:full_int}
\eeq
Evaluated analytically, the result can be written as two terms: ${\cal I}_{\alpha,\beta}^{(\nu)}(E_0)={\cal I}_{\alpha}^{(\nu)}(E_0)-{\cal I}_{\beta}^{(\nu)}(E_0)$, where
\begin{align*}
{\cal I}_\alpha^{(0)}(E_0)&={\cal I}_{\alpha,{\rm open}}^{(0)}+\kBT_\alpha\ln\left(1+e^{x_\alpha-y_\alpha}\right)\\
{\cal I}_\alpha^{(1)}(E_0)&={\cal I}_{\alpha,{\rm open}}^{(1)}\\
&+(\kBT_\alpha)^2\left[x_\alpha\ln\left(1+e^{x_\alpha-y_\alpha}\right)+{\rm Li}_2\left(-e^{x_\alpha-y_\alpha}\right)\right],
\end{align*}
with $x_\alpha=E_0/\kBT_\alpha$ and $y_\alpha=\mu_\alpha/\kBT_\alpha$, and the dilogarithm ${\rm Li}_2(x)=\sum_{k=1}^\infty x^k/k^2$.
Here, ${\cal I}_{\alpha}^{(\nu)}(E_0)$ and ${\cal I}_{\beta}^{(\nu)}(E_0)$ are just convenient abbreviations for expressions found when evaluating the \textit{full} Eq.~(\ref{eq:full_int}); they are \textit{not} to be understood as the result of Eq.~(\ref{eq:full_int}) with one Fermi function only, which would possibly not converge.
If the channel is fully open (if $E_0\rightarrow-\infty$), one recovers:
\begin{align*}
{\cal I}_{\alpha,{\rm open}}^{(0)}&=\mu_\alpha\\
{\cal I}_{\alpha,{\rm open}}^{(1)}&=-(\kBT_\alpha)^2\left[{\rm Li}_2\left(-e^{y_\alpha}\right){+}{\rm Li}_2\left(-e^{-y_\alpha}\right)\right].
\end{align*}
These analytical expressions have been employed to produce the data for plots shown in Figs.~\ref{fig:power_production}---\ref{fig:QPCQPC}.

\section{Linear response for fixed temperatures}
\label{app:linear}

Sec.~\ref{sec:tau_constraints} showed  that if $\tau_{\rm int}$ or $\tau_{\rm neq}$ are both electron-hole symmetric, then the demon conditions can generally not be fulfilled by tuning the  electrochemical potentials of the resource region, $\mu_3$ and $\mu_4$, only. Thus we study the constraints on the scatterers in a situation where tuning of $\mu_3$ and $\mu_4$ allows us to fulfill the demon conditions in this appendix.

\subsection{Three independent scatterers}
Hence, assuming that either $\tau_{\rm int}$ or $\tau_{\rm neq}$ are not electron-hole symmetric, we solve Eq.~\eqref{eq:strict_demon} and replace the obtained chemical potential affinities $\mathcal{A}_3^\mu$ and $\mathcal{A}_4^\mu$ in the expression for the current in terminal 2. We then get for the nonlocal contribution to the current
\begin{widetext}
\begin{align}
\label{eq:i2nl}
I_{2,{\rm nonloc}}=&\frac{1}{h}\left[\left(Y^{(1)}-Z^{(1)}\right)\mathcal{A}_3^T+Z^{(1)}\mathcal{A}_4^T\right]\nonumber\\
&+\frac{1}{h}\left\{\left[Y^{(0)}\left(h_{\rm neq}^{(1)}X_{\rm neq}^{(1)}-X_{\rm neq}^{(2)}h_{\rm neq}^{(0)}\right)-Z^{(0)}\left(g_{\rm int}^{(1)}X_{\rm neq}^{(1)}-X_{\rm neq}^{(2)}g_{\rm int}^{(0)}\right)\right]\mathcal{A}_3^T\right.\\
&\quad\quad\left.+\left[Y^{(0)}\left(\left(h_{\rm neq}^{(1)}\right)^2-h_{\rm neq}^{(2)}h_{\rm neq}^{(0)}\right)-Z^{(0)}\left(g_{\rm int}^{(1)}h_{\rm neq}^{(1)}-h_{\rm neq}^{(2)}g_{\rm int}^{(0)}\right)\right]\mathcal{A}_4^T\right\}\frac{1}{g_{\rm int}^{(1)}h_{\rm neq}^{(0)}-g_{\rm int}^{(0)}h_{\rm neq}^{(1)}}.\nonumber
\end{align}
\end{widetext}
 This only depends on the temperature bias in the resource region. The first term on the right hand side of Eq.~\eqref{eq:i2nl} is the direct response due to the temperature differences in the resource region. The second one derives from the chemical potential contribution under strict demon conditions.
Here we have defined $X_{i}^{(\nu)}=g_{\rm int}^{(\nu)}-h_{i}^{(\nu)}$, $Y^{(\nu)}=g_{\rm int}^{(\nu)}-h_{\rm ws}^{(\nu)}$ and $Z^{(\nu)}=h_{\rm neq}^{(\nu)}-q^{(\nu)}$.
Despite its lengthy expression, we see the requirements for a demonic effect. At least $\tau_\rmws$ or $\tau_{\rm int}$ must be energy dependent. Otherwise, $I_{2,{\rm nonloc}}=0$. This is surprising; it means that the working substance needs not be a thermoelectric conductor, so long as the interface is. Hence, the minimal requirement is that electron-hole symmetry is broken at the interface, even if $\tau_\rmneq$ and $\tau_\rmws$ are constant. In that case, we get
\beq
I_{2,{\rm nonloc}}=\frac{1-\tau_\rmws}{h}g_{\rm int}^{(1)}\left[(1-\tau_\rmneq)\mathcal{A}_3^T+\tau_\rmneq \mathcal{A}_4^T\right].
\eeq
Therefore terminal 4 can be disconnected, with the nonequilibrium distribution being composed by the mixing of $f_2$ and $f_3$ at the interface, only. This allows for a simpler experimental setup with only three terminals. However, it is of less interest, because the resource includes part of the working substance.

\subsection{Transparent interface}

Let us now consider the case where $\tau_{\rm int}=1$ such that $\tilde{f}(E)$ only depends on the resource reservoirs. Then the strict demon conditions read:
\begin{align}
g_0^{(0)}\mathcal{A}_{23}^\mu+g_\rmneq^{(0)}\mathcal{A}_{34}^\mu=g_\rmneq^{(1)}\left(\mathcal{A}_{4}^T-\mathcal{A}_3^T\right)\\
g_\rmneq^{(1)}\mathcal{A}_{34}^\mu=g_0^{(2)}\mathcal{A}_{3}^T+g_\rmneq^{(2)}\left(\mathcal{A}_{4}^T-\mathcal{A}_3^T\right),
\end{align}
where $g_0^{(\nu)}$ are obtained from Eq.~\eqref{eq:gin} with $\tau_i\equiv1$. A thermoelectric response of $\tau_\rmneq$ (i.e., having $g_\rmneq^{(1)}\neq0$)
allows there to be values of  $\mu_3$ and $\mu_4$ that satisfy the demon conditions for a given set of temperatures. This emphasizes the need for the nonequilibrium distribution to break electron-hole symmetry. This applies in the linear regime due to the symmetry of the energy derivative of the Fermi function in Eq.~\eqref{eq:hin}. As shown in Sec.~\ref{sec:parameters}, the energy dependence of $\tau_\rmneq$ is not required in the non-linear regime.  
The current can be written as
\begin{align}
&I_{2,{\rm nonloc}}=\frac{1}{h}Z^{(1)}\mathcal{A}_{43}^T\\
&{-}\frac{1/h}{g_0^{(0)}g_\rmneq^{(1)}}\left\{\left[\left(g_\rmneq^{(1)}\right)^2Y^{(0)}{+}g_\rmneq^{(2)}R\right]\mathcal{A}_{43}^T+g_0^{(2)}R\mathcal{A}_3^T\right\},\nonumber
\end{align}
where $R=g_\rmws^{(0)}g_\rmneq^{(0)}-g_0^{(0)}q^{(0)}$. Here a constant $\tau_\rmws$ makes $R=0$, $Y^{(0)}=(1-\tau_\rmws)g_0^{(0)}$ and $Z^{(1)}=(1-\tau_\rmws)g_\rmneq^{(1)}$, resulting in $I_{2,{\rm nonloc}}=0$. This means that, when the interface is transparent, the resource and the working substance must be thermoelectric systems separately. 

Note that the opposite is not true: a $\tau_{\rmws}$ that is energy-dependent but does not break electron-hole symmetry gives a finite ``demonic" response, in general ---but no conventional thermoelectric effect. This is the case, for instance, of a quantum dot whose resonance is aligned with $\mu_1$, as discussed in Sec.~\ref{sec:parameters}.

\bibliography{cite.bib}

\begin{thebibliography}{113}
\bibitem [{ing()}]{ingeneral}%
  \BibitemOpen
  \href@noop {} {}\bibinfo {note} {In general $I^E_i$ depends on the gauge
  (depends on the choice of the zero of energy), however $I^E_3+I^E_4$ does not
  if $I_3+I_4=0$.}\BibitemShut {Stop}%
\bibitem [{tho()}]{thot}%
  \BibitemOpen
  \href@noop {} {}\bibinfo {note} {We have $T_\text{hot}$ in $F_\text{neq}$
  here for the following reason (cf. \cite{temperature}). If reservoirs 3 and 4
  were decoupled from the nonequilibrium resource, the refrigerator would cease
  to function, and so both the nonequilibrium resource and the cold reservoir
  would eventually thermalize at $T_\text{hot}$.}\BibitemShut {Stop}%
\end{thebibliography}

\end{document}